\documentclass[twocolumn, twocolappendix]{aastex631}
\usepackage{amsmath}

\received{July 16, 2024}
\revised{September 25, 2024}
\accepted{September 28, 2024}
\submitjournal{ApJ}
\shorttitle{Effect of Viscosity in the ICM}
\shortauthors{Marin-Gilabert et al.}

\begin{document}

\title{Density Fluctuations in the Intracluster Medium: An Attempt to Constrain Viscosity with Cosmological Simulations}

\email{tmarin@usm.lmu.de}

\author[0000-0001-8390-9356]{Tirso Marin-Gilabert}
\affiliation{Universitäts-Sternwarte, Fakultät für Physik, Ludwig-Maximilians-Universität München, Scheinerstr.1, 81679 München, Germany}

\author[0000-0001-8867-5026]{Ulrich P. Steinwandel}
\affiliation{Center for Computational Astrophysics, Flatiron Institute, 162 5th Avenue, New York, NY 10010}

\author[0000-0002-0796-8132]{Milena Valentini}
\affiliation{Astronomy Unit, Department of Physics, University of Trieste, via Tiepolo 11, I-34131 Trieste, Italy}
\affiliation{INAF - Osservatorio Astronomico di Trieste, via Tiepolo 11, I-34131 Trieste, Italy}
\affiliation{INFN, Instituto Nazionale di Fisica Nucleare, Via Valerio 2, I-34127, Trieste, Italy}
\affiliation{ICSC - Italian Research Center on High Performance Computing, Big Data and Quantum Computing, via Magnanelli 2, 40033, Casalecchio di Reno, Italy}

\author[0000-0003-2656-5985]{David Vall\'{e}s-P\'{e}rez}
\affiliation{Departament d'Astronomia i Astrof\'{\i}sica, Universitat de Val\`encia, C/Doctor Moliner 50, E-46100 Burjassot (Val\`encia), Spain}

\author{Klaus Dolag}
\affiliation{Universitäts-Sternwarte, Fakultät für Physik, Ludwig-Maximilians-Universität München, Scheinerstr.1, 81679 München, Germany}
\affiliation{Max-Planck-Institut für Astrophysik, Karl-Schwarzschild-Straße 1, 85741 Garching, Germany}

\begin{abstract}

The impact of viscosity in the Intracluster Medium (ICM) is still an open question in astrophysics. To address this problem, we have run a set of cosmological simulations of three galaxy clusters with a mass larger than $M_{\mathrm{Vir}} > 10^{15} $M$_{\odot}$ at $z=0$ using the SPMHD-code \textsc{OpenGadget3}. We aim to quantify the influence of viscosity and constrain its value in the ICM. Our results show significant morphological differences at small scales, temperature variations, and density fluctuations induced by viscosity. We observe a suppression of instabilities at small scales, resulting in a more filamentary structure and a larger amount of small structures due to the lack of mixing with the medium. The conversion of kinetic to internal energy leads to an increase of the virial temperature of the cluster of $\sim$5\% - 10\%, while the denser regions remain cold. The amplitude of density and velocity fluctuations are found to increase with viscosity. However, comparison with observational data indicates that the simulations, regardless of the viscosity, match the observed slope of the amplitude of density fluctuations, challenging the direct constraint of viscosity solely through density fluctuations. Furthermore, the ratio of density to velocity fluctuations remains close to 1 regardless of the amount of viscosity, in agreement with the theoretical expectations. Our results show for the first time in a cosmological simulation of a galaxy cluster the effect of viscosity in the ICM, a study that is currently missing in the literature.

\end{abstract}


\section{Introduction} \label{sec:intro}

The Intracluster Medium (ICM) is the dominant baryonic component visible in galaxy clusters, filling the gravitational potential with hot ($T \sim 10^7-10^8$ K) ionised, X-ray emitting gas. This gas is continually perturbed by galaxy motions \citep[e.g.][]{Faltenbacher_2005}, mergers \citep[e.g.][]{ZuHone_2011, Iapichino_2017}, AGN outflows \citep[e.g.][]{O'Neill_2009, Gaspari_2015} and accretion of gas along filaments \citep[e.g.][]{Kravtsov_2012, Valles-Perez_2021b}. These processes inject energy at large scales (low wavenumber $k$), which decays towards higher $k$ modes in a Kolmogorov-like cascade, introducing turbulence on a wide range of scales \citep[e.g.][]{Kolmogorov_1941, Kolmogorov_1962}. The energy is then dissipated into heat, affecting small scale processes such as cosmic ray re-acceleration \citep[e.g.][]{Fujita_2003, Brunetti_2007}, star formation \citep[e.g.][]{Federrath_2016} or magnetic field amplification \citep[e.g.][]{Kazantsev_1968, Kulsrud_1992}. These complex mechanisms over cosmological periods of time produce very complicated scenarios in which the transport properties of the gas are fundamental. 

Probing the turbulent nature of the ICM via observations is crucial for understanding the gas dynamics and, therefore, the physics and gas properties of the cluster \citep{Gilfanov_1987, Churazov_2004}. The Hitomi collaboration \citep{Hitomi_2018} measured subsonic speeds of the gas with a 3D mach number of $\mathcal{M}_{\mathrm{3D}} \sim 0.3-0.45$ in Perseus. However, with the exception of the Hitomi collaboration, direct measurements of velocities are not available yet (although they soon will be with the launch of \textit{XRISM} in 2023, a high-resolution X-ray spectrometer, \citealp{XRISM_2020}). Therefore, indirect measurements are needed to infer the dynamics of the gas. This can be done by measuring the X-ray surface brightness and pressure fluctuations \citep{Soltan_1990, Schuecker_2004, Churazov_2012} and linking them to density fluctuations induced by turbulence. \citet{Zhuravleva_2014} showed theoretically that subsonic gas motions driven on large scales in a stratified atmosphere can introduce density fluctuations. The amplitude of these density fluctuations is expected to scale linearly with the amplitude of the one-component velocity fluctuations at each scale \citep{Zhuravleva_2014}. 

This relation has been tested in galaxy cluster simulations, where \citet{Zhuravleva_2014} found a proportionality coefficient between density and velocity amplitude fluctuations of $\eta_{\rho} \approx 1.0\pm0.3$, as predicted theoretically, and \citet{Gaspari_2014} a value of $\eta_{\rho} \approx 1.3$. Using cosmological simulations of galaxy cluster formation, \citet{Simonte_2022} found that this relation depends on the dynamical state of the cluster, with $\eta_{\rho} \approx 1.15\pm0.06$ for relaxed clusters, but a flatter relation and a larger dispersion for unrelaxed clusters persists. The importance of the dynamical state was tested in \citet{Zhuravleva_2023}, showing that relaxed clusters tend to have a proportionality coefficient closer to 1 than unrelaxed clusters, where the coefficient tends to be larger. This is because in unrelaxed clusters, the assumption of a nearly hydrostatic atmosphere does not hold anymore, producing a larger scatter in the density-velocity fluctuations. \citet{Zhuravleva_2023} also showed that accounting for halo ellipticity might be important, especially for the inner regions of relaxed clusters. The density-velocity fluctuations relation is not universal, but depends on the level of stratification, characterized by the Richardson number \textit{Ri}\footnote{On large scales, stratification is expected to be dominant over turbulence ($Ri > 1$), while on small scales turbulence dominates ($Ri < 1$) \citep{Mohapatra_2020}.}. \citet{Mohapatra_2020} found that $\eta_{\rho}$ increases with \textit{Ri}, ranging from $0.01 \lesssim \eta_{\rho}^2 \lesssim 1$ for \textit{Ri} from 0 to 10.

The properties of these density perturbations do not only reflect the driving mechanisms that trigger them, but also depend on the microphysics of the ICM, specifically, for thermal conduction \citep{Ruszkowski_2011} and viscosity \citep{Zhuravleva_2019, Kunz_2022}. \citet{Zhuravleva_2019} suggested to use observations of density fluctuations to constrain the amount of viscosity in galaxy clusters. The idea is to compare the amplitude of density fluctuations measured in observations against simulations with different amounts of viscosity. Using Coma observations and simulations of incompressible hydrodynamic turbulence carried out using Direct Numerical Simulations (DNS), \citet{Zhuravleva_2019} concluded that the effective isotropic viscosity must be suppressed by at least a factor of $\sim$10 to $\sim$1000 with respect to the Spitzer value. This suppression could be the effect of magnetic fields, which reduce the effect of viscosity \citep{Zuhone_2015, Berlok_2019, Squire_2023}, as well as an enhanced scattering rate due to plasma instabilities \citep{Schekochihin_2006, Kunz_2014, Berlok_2021}.

The physics of the ICM has been studied extensively using simulations of isolated systems with more idealised setups \citep[e.g.][]{ZuHone_2009, Bruggen_2012, Mohapatra_2020, Zhang_2020} and cosmological simulations \citep[e.g.][]{Dolag_2005, Teyssier_2011, Vazza_2016, Pakmor_2023, Steinwandel_2024}. Similarly, viscosity in the context of ICM has also been extensively studied with simulations, focusing on its effect on AGN-powered bubbles \citep[][]{Scannapieco_2008, Dong_2009}, on the suppression of instabilities in cold fronts \citep{Roediger_2011, Suzuki_2013, Zuhone_2015} and on its impact on stripped galaxies \citep{Roediger_2015, Kraft_2017}. 

However, viscosity in the intracluster medium (ICM) has not been extensively studied using cosmological simulations. The only study in the literature, by \citet{Sijacki_2006}, did not provide a detailed analysis of these effects. In our study, we conduct a set of cosmological simulations focusing on how viscosity affects massive clusters in a realistic scenario. We use different values of the isotropic Spitzer viscosity \citep{Spitzer_1962, Braginskii_1965} and compare it with the non-viscous case. Our goal is to provide a general view of how the morphology of the cluster is affected by viscosity and how the virial temperature depends on the amount of viscosity. We also want to test the criteria to constrain viscosity suggested by \citet{Zhuravleva_2019} and if this criterion holds in the context of the complex, realistic cosmological simulation. To this end, we compare our results with observations of density fluctuations from \citet{Heinrich_2024}. Additionally, we want to verify if the density fluctuations scale linearly with the velocity fluctuations, independently of the how viscous the ICM is. Although full Spitzer viscosity is larger than expected, due to the different mechanisms that suppress the viscous stress, we want to show this extreme case for a better understanding of the ongoing mechanisms. In a follow up work, we will show additional consequences of a viscous ICM such as changes of the magnetic field amplification, effects in the merger history or the different energy distribution, together with the time evolution of gas properties.

This paper is organised as follows. In Sec.~\ref{sec:methods} we introduce the relevant equations and our setup. In Sec.~\ref{sec:results} we show the results obtained from our simulations and the comparison with observations. Finally, we discuss the results obtained and conclude in Sec.~\ref{sec:conclusions}.

\section{Methods} \label{sec:methods}

\subsection{Theoretical considerations} \label{sec:theory}

The ICM can be understood as a compressible magnetised plasma, therefore it can be described using the equations of magnetohydrodynamics (MHD):
\begin{equation}
    \frac{\mathrm{d} \rho}{\mathrm{d} t} + \rho \nabla \cdot \mathbf{v} = 0 ,
\end{equation}
\begin{equation}
    \rho \frac{\mathrm{d} \mathbf{v}}{\mathrm{d} t} + \nabla P = -\rho \nabla \Phi - \nabla \cdot \boldsymbol{\Pi} + \frac{(\nabla \times \mathbf{B}) \times \mathbf{B}}{4\pi} ,
\end{equation}
\begin{multline}
    \frac{\mathrm{d} E}{\mathrm{d} t} + \mathbf{v} \cdot \nabla P + (E + P)\nabla \cdot \mathbf{v} - \nabla \cdot \frac{\mathbf{B} (\mathbf{v} \cdot \mathbf{B})}{4 \pi} = \\ -\rho \mathbf{v} \cdot \nabla \Phi - \nabla \cdot \mathbf{Q} - \nabla \cdot \left( \boldsymbol{\Pi} \cdot \mathbf{v} \right)
\end{multline}
\begin{equation}
    \frac{\partial \mathbf{B}}{\partial t} = \nabla \times \left(\mathbf{v} \times \mathbf{B} \right)  ,
\end{equation}
where
\begin{equation}
    \frac{\mathrm{d}}{\mathrm{d}t} = \frac{\partial}{\partial t} + (\mathbf{v} \cdot \nabla)
\end{equation}
is the Lagrangian derivative. $\rho$ is the gas density, $\mathbf{v}$ the velocity of the fluid, $\mathbf{B}$ the magnetic field, $\mathbf{Q}$ is the heat flux and $\Phi$ is the gravitational potential. $E$ is the energy per unit volume (kinetic + thermal + magnetic)
\begin{equation}
    E = \frac{\rho \mathbf{v}^2}{2} + \rho u + \frac{\mathbf{B}^2}{8 \pi} \, ,
\end{equation}
with $u$ being the specific internal energy. $P$ is the pressure
\begin{equation}
    P = (\gamma - 1) \rho u \, ,
\end{equation}
with an adiabatic index of $\gamma = 5/3$ for monoatomic gases. $\boldsymbol{\Pi}$ is the viscous stress tensor, which defines the behaviour of Spitzer viscosity as
\begin{equation}
   \boldsymbol{\Pi} = \eta \, \sigma_{ij} = \eta \, \left( \frac{\partial v_i}{\partial x_j} + \frac{\partial v_j}{\partial x_i} - \frac{2}{3} \delta_{ij} \frac{\partial v_k}{\partial x_k} \right) \, ,
    \label{eqn:viscous_stress_tensor}
\end{equation}
where we have dropped the bulk viscosity term, since it is related to the degrees of freedom of molecular rotations, being zero for monoatomic gases \cite[see e.g][]{Zeldovich_1967, Pitaevskii_1981}. The viscous stress tensor is equal to the rate of strain tensor ($\sigma_{ij}$) multiplied by the shear viscosity coefficient $\eta$, defined as 
\begin{equation}
    \eta = 0.406 \frac{m_i^{1/2}(k_B T_i)^{5/2}}{(Z\,e)^4 \ln{\Lambda}} \, ,
    \label{eqn:viscosity}
\end{equation} 
where $m_i$ is the mass of the proton, $T_i$ is the temperature of the plasma, $Ze$ is the ion charge and $\ln \Lambda$ is the Coulomb logarithm.

Assuming subsonic gas motions within the galaxy cluster ($v \ll c_s$) and a stratified atmosphere, density perturbations are expected to be proportional to the one-component velocity at each scale $l = 1/k$
\begin{equation}
    \left( \frac{\delta \rho}{\rho} \right)_k^2 = \eta_{\rho}^2 \left(\frac{v_{1\mathrm{D}}}{c_s} \right)_k^2 = \eta_{\rho}^2 \, \mathcal{M}_{\mathrm{1D},k}^2 \, .
    \label{eqn:eta_relation}
\end{equation}
$c_s$ is the soundspeed of the medium and $\mathcal{M}_{\mathrm{1D},k}$ the 1D mach number. A complete derivation can be found in \citet{Zhuravleva_2014}.

\subsection{Viscosity saturation} \label{sec:viscosity_saturation}

A proper treatment of viscosity requires the implementation of a viscosity saturation to avoid unphysical results. The viscous stress saturates when the scale over which the velocity varies becomes smaller than the ion mean free path. This means that the viscous momentum transfer propagates faster than the mean soundspeed of the medium, overestimating this momentum transfer \citep{Sarazin_1986}. 

To avoid this, we need to introduce a viscosity saturation that limits the momentum propagation and ensures a smooth transition from the non-saturated to the saturated state. We do this in such a way that the momentum transfer of the saturated viscosity propagates at a velocity comparable to the soundspeed of the medium. We define the velocity length scale as $l_v = 2\,c_s / |\sigma_{ij}|$, where 
\begin{multline}
    |\sigma_{ij}| = \sqrt{\mathrm{tr}(\sigma_{ij}^2)} = \sqrt{\mathrm{tr}(\sigma_{ij} \cdot \sigma_{ij})} =\\ \sqrt{\mathrm{tr}(\sigma_{xx}^2 + \sigma_{yy}^2 + \sigma_{zz}^2 + 2\sigma_{xy}^2 + 2\sigma_{xz}^2 + 2\sigma_{yz}^2)}
\end{multline}
is the strength of the rate of strain tensor of the viscous stress tensor. We now introduce the viscosity saturation as the analogous to the thermal conduction saturation \citep{Cowie_1977}, following the implementation done in \cite{Su_2017}:
\begin{equation}
    \eta_{\mathrm{Sat}} = \frac{\eta}{1 + 4.2 \frac{\lambda_I}{l_v}} \, ,
    \label{eqn:saturation}
\end{equation}
where $\lambda_I$ is the ion mean free path:
\begin{equation}
    \lambda_I = \frac{3^{3/2} (k_B T_I)^2}{4\pi^{1/2} n_I e^4 \ln{\Lambda}} \, .
\end{equation}
Here, $T_I$ is the ion temperature, $n_I$ the ion number density, $e$ the electric charge and $\ln{\Lambda} = 37.8$ the Coulomb logarithm. This way we make sure that when $l_v < \lambda_I$, the shear viscosity coefficient saturates, avoiding unphysical large values of the viscous stress tensor $\boldsymbol{\Pi}$ and numerical issues like extremely small timesteps.

\subsection{Simulation Setup}

We perform cosmological, magneto-hydrodynamical simulations of galaxy clusters using the smoothed particle magneto-hydrodynamics (SPMHD) code \textsc{OpenGadget3} \citep{Springel_2005, Groth_2023}. Gravity is solved via the Tree-PM method, where the long-distance gravitational forces are computed on a PM mesh and the short-distance forces are computed on a gravity tree. For the hydro computation we used the modern SPH implementation \citep{Beck_2015}, including artificial viscosity \citep{Balsara_1995, Cullen_2010} and artificial conductivity \citep{Price_2008}, with a Wendland $C^6$ kernel \citep{Wendland_1995, Dehnen_2012} and 295 neighbours. This is necessary to avoid the `pairing instability' \citep{Price_2012} and the `E$_0$ error' \citep{Read_2010} and capture mixing properly \citep{Tricco_Price_2013, Hu_2014}.

We also include magnetic fields based on the implementation of \cite{Bonafede_2011} and \cite{Stasyszyn_2013}, with an initial seed of $B_{\mathrm{seed}} = 10^{-12}$ G (comoving) in the $x$-direction, which corresponds to an initial magnetic field of $B_{\mathrm{ini, ph}} = B_{\mathrm{seed}} \cdot (1 + z_{\mathrm{ini}})^2 = 1.98\times10^{-8}$ G in physical units. The choice of this initial magnetic field leads to a saturation of the dynamo at z$\sim$1.5 at this resolution \citep[][Marin-Gilabert et al, in prep]{Steinwandel_2022}. Our simulations also include anisotropic thermal conductivity via a bi-conjugate gradient solver \citep{Arth_2017, Steinwandel_2022}. 

To properly understand the effects of physical viscosity, we perform three different simulations for each cluster: one with no viscosity (labeled as ``Ideal"), one with $1/3$ of Spitzer viscosity (labeled as ``$1/3 \, \eta$") and one with full Spitzer viscosity (labeled as ``$\eta$"). The implementation of viscosity in \textsc{OpenGadget3} is described in \cite{Sijacki_2006} and \cite{Marin-Gilabert_2022}. Now, we additionally include a viscous saturation (see section \ref{sec:viscosity_saturation} for details) to avoid unphysical results \citep{Cowie_1977, Sarazin_1986, Su_2017}. On top of the physical viscosity, a higher order shock capturing method (i.e. artificial viscosity) is necessary to properly capture shocks in SPH \citep{Monaghan_1983, Monaghan_1992} even when physical viscosity is implemented \citep{Sijacki_2006}.

We want to focus on understanding the complex properties of gas in the ICM, in particular the effect of viscosity and its observational implications. For this reason, we run non-radiative simulations without the effects of subgrid models (e.g. star formation and feedback).

\subsection{Initial Conditions}

We ran a total of nine zoom-in simulations of three different galaxy clusters from the Dianoga suite \citep{Bonafede_2011, Ragone_Figueroa_2013}: the g5503149, g1657050 and g6348555 regions (labeled as g55, g16 and g63, respectively) at the 10x resolution level. These clusters were taken from a low-resolution $N$-body cosmological simulation of a periodic box of 1 $h^{-1}$ Gpc comoving size. Each Lagrangian region was then re-simulated with a higher resolution using the \textit{zoomed initial conditions technique} \citep{Tormen_1997}. 

We chose the g55, g16 and g63 clusters because they have a $M_{\mathrm{Vir}}$\footnote{The suffix \textit{Vir} indicates the virialized gas, i.e. the gas within the galaxy cluster that satisfies the virial theorem.} $> 10^{15} $M$_{\odot}$ at $z=0$; g55 and g16 are expected to be unrelaxed, while g63 is very relaxed. To classify the clusters as relaxed or unrelaxed, we used the centre-of-mass offset, where we measured the separation between the density peak position and the center of mass within $R_{\mathrm{200c}}$ \citep{Power_2012, Cui_2016}. The adopted cosmological parameters are $\Omega_\mathrm{0}= 0.24$, $\Omega_\mathrm{\Lambda}= 0.76$, $\Omega_\mathrm{b}= 0.04$, $h = 0.72$ and $\sigma_8 = 0.8$. The initial redshift for all the simulations is $z_{\mathrm{ini}} = 140$ and the mass resolution is $m_{\mathrm{gas}} = 1.56\times10^{7} $M$_{\odot}$ in gas particles and $m_{\mathrm{dm}} = 8.44\times10^{7} $M$_{\odot}$ in dark matter. The choice of this particle resolution is due to the convergence in the magnetic field amplification shown in \citet{Steinwandel_2022} against a lower particle resolution. The details of each cluster at $z=0$ are shown in table \ref{tab:clusters}.

\begin{table*}[]
\centering
\caption{Properties of the three different clusters used in this work with three different amounts of viscosity. First column shows the virial mass, second column the virial temperature, third column the $R_{\mathrm{2500c}}$ and fourth column shows $R_{\mathrm{200c}}$, each at $z=0$.}
\begin{tabular}{cccc|ccc|ccc|ccc}
    & \multicolumn{3}{c|}{M$_{\mathrm{Vir}}$ {[}$10^{15}$  M$_{\odot}${]}}       & \multicolumn{3}{c|}{T$_{\mathrm{Vir}}$ {[}K{]}}        & \multicolumn{3}{c|}{$R_{\mathrm{2500c}}$ {[}kpc{]}} & \multicolumn{3}{c}{$R_{\mathrm{200c}}$ {[}kpc{]}}          \\ \cline{2-13}
    & Ideal               & 1/3 $\eta$          & $\eta$              & Ideal            & 1/3 $\eta$       & $\eta$           & Ideal          & 1/3 $\eta$         & $\eta$        & Ideal          & 1/3 $\eta$        & $\eta$       \\ \hline \hline
\multicolumn{1}{c|}{g55} & $1.34$ & $1.38$ & $1.33$ & $7.05\times10^7$ & $7.32\times10^7$ & $7.39\times10^7$ & 498.21         & 509.77             & 510.40        & 3000.96        & 3020.54           & 2995.14      \\ \hline
\multicolumn{1}{c|}{g16} & $1.44$ & $1.40$ & $1.20$ & $6.22\times10^7$ & $6.29\times10^7$ & $6.22\times10^7$ & 459.07         & 460.04             & 435.29        & 3232.38        & 3204.37           & 3086.66   \\ \hline
\multicolumn{1}{c|}{g63} & $1.11$ & $1.22$ & $1.17$                    & $6.40\times10^7$ & $7.81\times10^7$ & $6.91\times10^7$                 & 455.19         & 499.39             & 471.84              & 2939.73        & 3019.53        & 2976.75          
\end{tabular}
\label{tab:clusters}
\end{table*}

\section{Results} \label{sec:results}

In this section we present the results obtained from the three different clusters and the three different values of viscosity employed. We focus first on the intrinsic effect of viscosity with respect to the non-viscous case and then we will try to constrain the viscosity from X-ray observations.

\subsection{Morphology} \label{sec:morphology}

\begin{figure*}
    \centering
	\includegraphics[width=\textwidth]{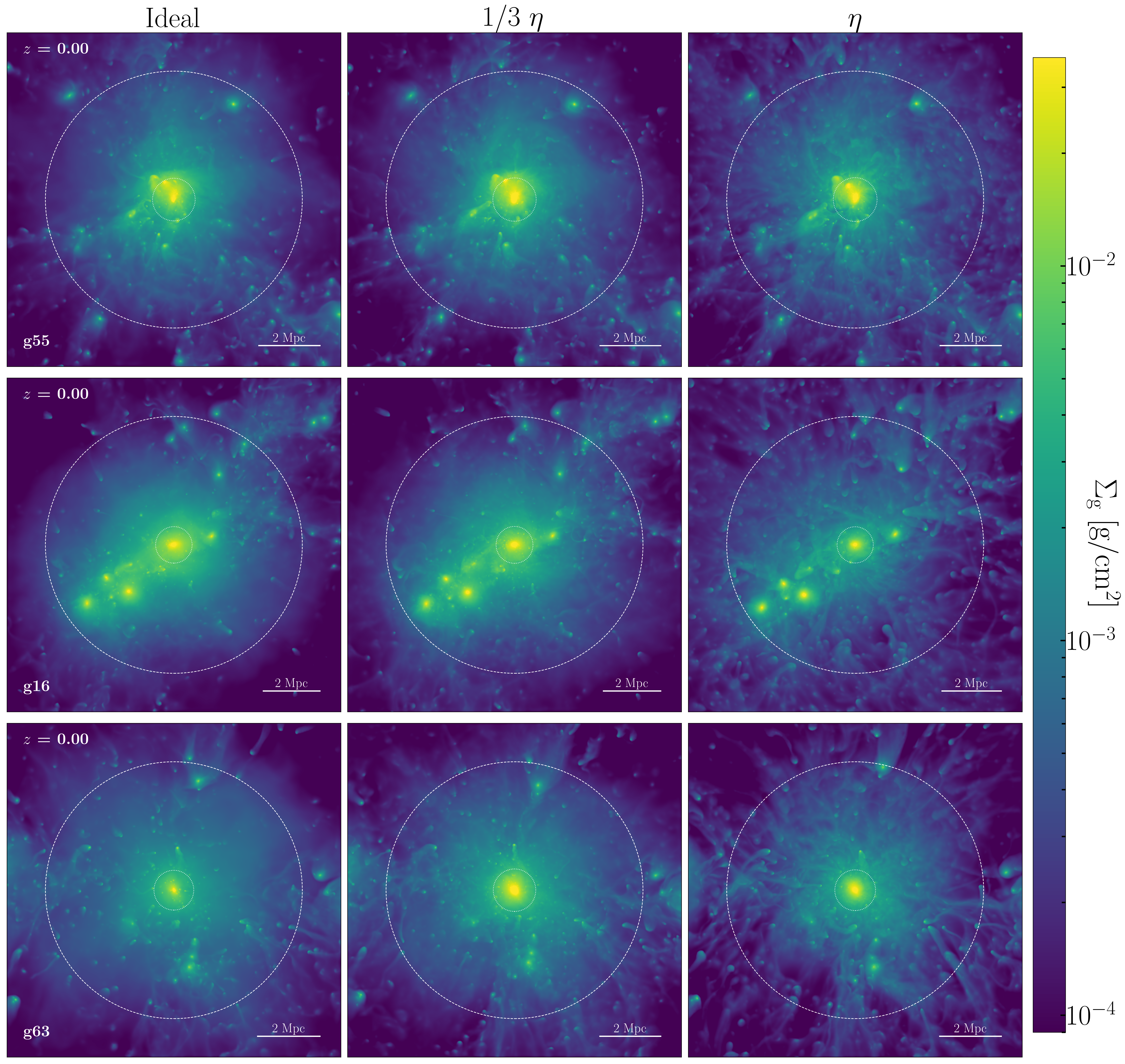}
    \caption{Density projections for all of our simulations listed in table \ref{tab:clusters}. From left to right: MHD only, MHD with $1/3$ of Spitzer viscosity and full Spitzer viscosity at redshift zero. From top to bottom: the runs g55, g16 and g63.}
    \label{fig:colormaps_density}
\end{figure*}

Viscosity acts by transforming kinetic energy into internal energy, leading to a suppression of the growth of hydrodynamical instabilities and to an increase of the gas temperature \citep{Roediger_2013, Marin-Gilabert_2022}. This is expected to produce a strong effect in the morphology of galaxy clusters at small scales (few kpc), where the turbulent cascade is truncated due to the effect of viscosity. Fig.~\ref{fig:colormaps_density} shows the surface density for each of the three clusters and the three different configurations. The inner dotted circle corresponds to $R_{\mathrm{2500c}}$ and the outer dashed circle corresponds to $R_{\mathrm{200c}}$\footnote{In this paper, $R_{\mathrm{2500c}}$ ($R_{\mathrm{200c}}$) is defined as the radius enclosing the region of the cluster with a mean density 2500 (200) times larger than the critical density of the universe.}. 

By comparing each cluster independently, we can observe that there are no major changes at large scales (few hundred kpc - Mpc), where the big structures remain in the same position in the three different cases and the overall gas distribution is very similar. However, the ICM is more homogeneous in the non-viscous case, whereas in the viscous cases we find sharper density discontinuities and more regions with very diffuse gas.

Although at large scale the picture is similar, we find many differences at smaller scales. The lack of mixing due to viscosity leads to more small structures that have not been disrupted due to the growth of instabilities. The survival of the clump will depend on different parameters such as the size of the clump, its velocity or the overdensity with respect to the medium \citep{Klein_1994, Pittard_2005, Valentini_Brighenti_2015}, which is beyond of the scope of this paper. The growth or suppression of instabilities will also depend on the amount of viscosity, which depends on the temperature, generating a very complex system difficult to analyse in cosmological simulations. However, its effect can be seen in the amount of clumps observed in Fig.~\ref{fig:colormaps_density}. 

Another morphological difference observed in the cases with viscosity is the filamentary structure. This is produced by the infalling structures moving towards the center of the cluster, which experience a ram pressure stripping \citep[e.g.][]{Gisler_1976, Nulsen_1982, Randall_2008}. This stripped gas produces a tail with a density contact discontinuity with respect to the medium, resulting in the growth of instabilities. Viscosity slows down the growth process of these instabilities, leading to longer tails that last for longer periods of time \citep{Roediger_2015, Kraft_2017}. Although magnetic fields can also produce these filamentary structures \citep{Das_2023}, all our simulations include magnetic fields, excluding it as the origin of these differences.

\begin{figure*}
    \centering
	\includegraphics[width=\textwidth]{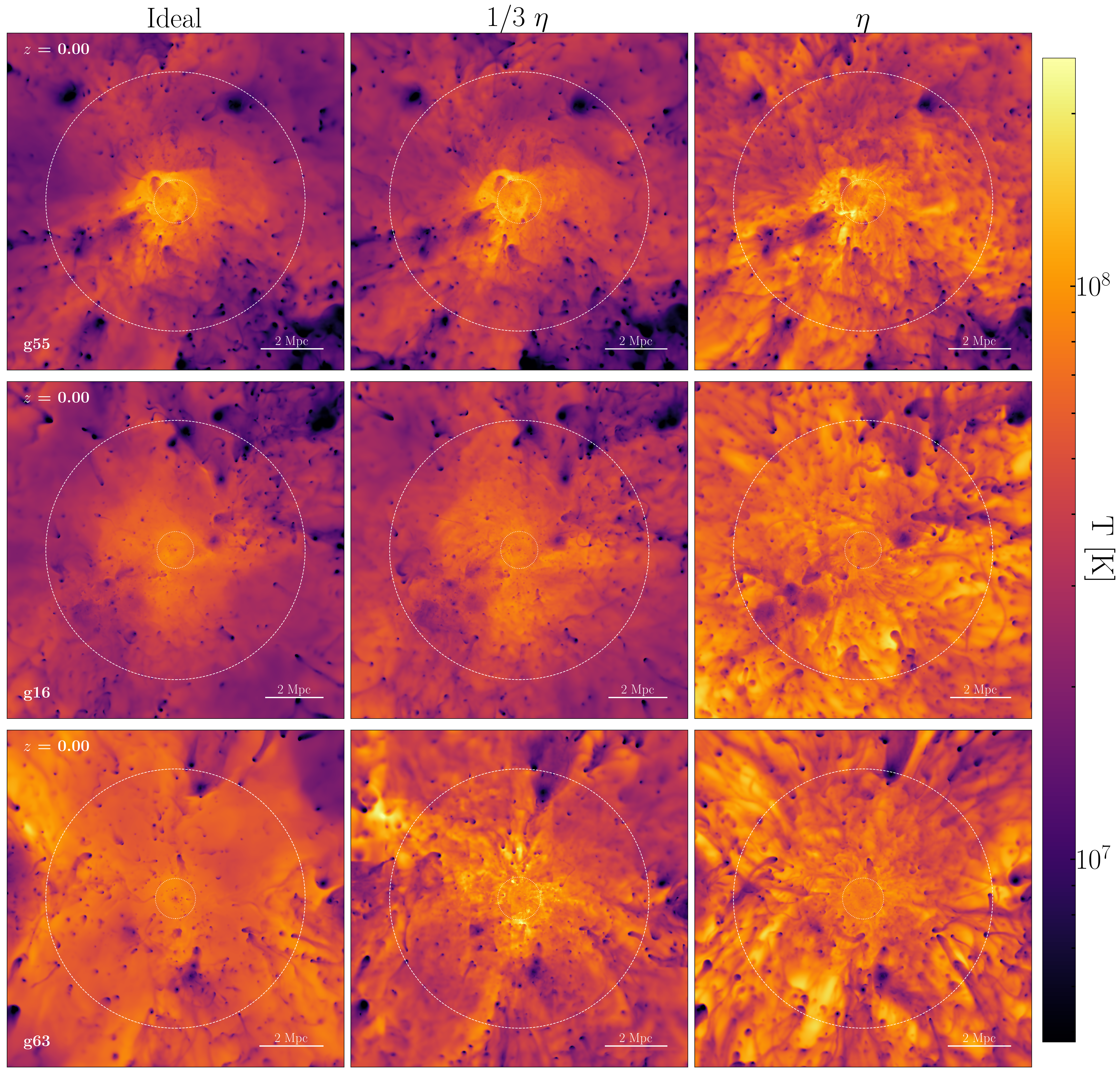}
    \caption{Temperature projections for all of our simulations listed in table \ref{tab:clusters}. From left to right: MHD only, MHD with $1/3$ of Spitzer viscosity and full Spitzer viscosity at redshift zero. From top to bottom: the runs g55, g16 and g63.}
    \label{fig:colormaps_temperature}
\end{figure*}

Because of the  strong dependence of viscosity with temperature one might expect that the above-mentioned differences might correlate well with the temperature of the cluster. In Fig.~\ref{fig:colormaps_temperature} we show temperature colormaps of the nine different simulations. The cases with more viscosity lead to higher temperatures in the diffuse gas, however, the temperatures of the dense clumps are very similar in all three cases. Overall, the virial temperature at $z=0$ is around 5\% to 10\% higher in the viscous cases compared to the ideal ones due to heat dissipation (see table \ref{tab:clusters}). We find again a more homogeneous temperature in the ideal runs, while more fluctuations are present in the viscous runs. Due to the dependence of viscosity on temperature, viscosity starts to become important at low redshift, when the virial temperature becomes larger than $\sim 10^7$~K. In this respect, the merger history of the cluster becomes very important, as mergers behave differently whether we have or we do not have viscosity. This is analysed in full depth in a follow up work. The hotter medium leads to a larger viscosity, which heats up the medium, producing a larger viscosity, entering a runaway cycle which is prevented by the viscous saturation, avoiding unphysical temperatures (see section \ref{sec:viscosity_saturation}). 

\subsection{Density and velocity fluctuations} \label{sec:fluctuations}
\subsubsection{Gas distribution} \label{sec:gas_distribution}

Quantifying the density fluctuations observed in Fig.~\ref{fig:colormaps_density} is fundamental for understanding the small scale processes which determine the ICM properties. Comparing with observations can help us to constrain the amount of viscosity in the ICM, as suggested in \citet{Zhuravleva_2019}. To this end, we first remove the high-density clumps, leaving only the bulk gas (source of the X-ray emission in the ICM), and then compute the density and velocity fluctuations.

\begin{figure}
	\includegraphics[width=\columnwidth]{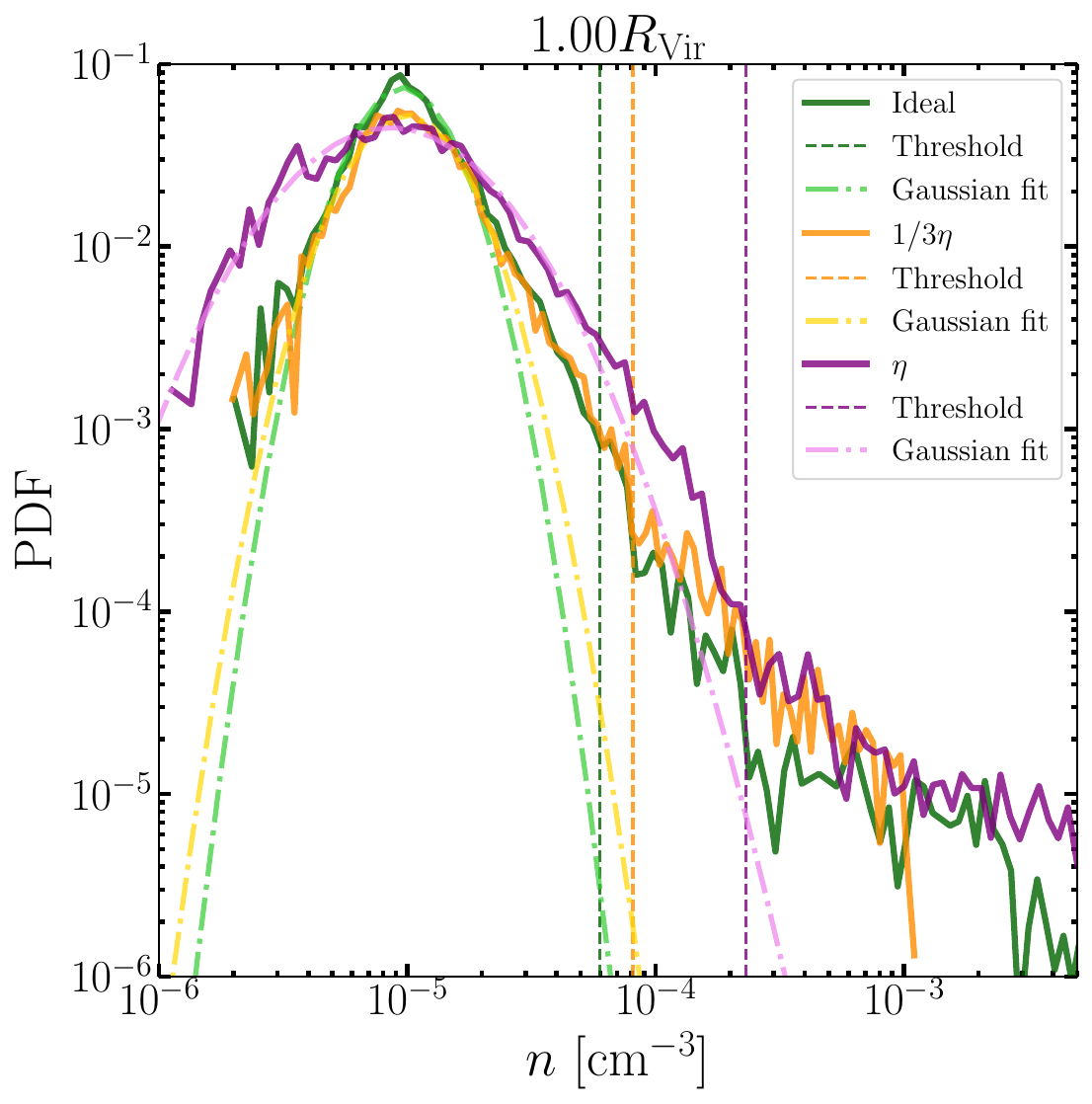}
    \caption{Density probability density function (PDF) of a thin shell of 5~kpc of the cluster g55 to show the effect of viscosity on the gas density distribution. The solid lines show the data; the vertical dashed lines indicate the values of the density thresholds separating bulk and high-density gas; and the dash-dotted lines the log-normal distribution fit to the bulk density component of the gas.}
    \label{fig:pdf_density}
\end{figure}

To remove the high-density regions we follow the method introduced in \cite{Zhuravleva_2012}: we divide the galaxy cluster into spherical shells and compute the density PDF for each of the shells. We calculate the median value of each shell and, alongside it, a threshold value to separate the bulk gas from the high-density gas. A particle is considered to belong to the high-density gas if the following criterion is met: 
\begin{equation}
    \log_{10} n > \log_{10} \{n\} + f_{\mathrm{cut}} \sigma_{10} \, ,
    \label{eqn:threshold}
\end{equation}
where $\{n\}$ is the median value of the density in the shell, $f_{\mathrm{cut}}$ is a parameter tuned to select more or less gas as high-density and $\sigma_{10}$ is the standard deviation ($\log_{10}$ based) assuming a log-normal distribution of the density PDF. $\sigma_{10}$ can be expressed as
\begin{equation}
    \sigma_{10} = \frac{W_{10}}{2\sqrt{2\ln 2}} \, ,
    \label{eqn:sigma10}
\end{equation}
with
\begin{equation}
    W_{10} (n) = \log_{10}\frac{n_2}{n_1} \, .
    \label{eqn:W10}
\end{equation}
$W_{10}$ accounts for the logarithmic interval where 76\% of the particles are contained, $n_1$ corresponds to the 12th percentile of density and $n_2$ to the 88th percentile. The value of $f_{\mathrm{cut}}$ is set to be between 2.5 and 4.5 \citep{Zhuravleva_2012}. A lower value will displace the threshold towards lower densities, selecting a larger number of particles as high-density gas, while a higher value will displace it towards larger densities. We choose a value of $f_{\mathrm{cut}}=2.5$ for our analysis. The reason for this choice can be seen in Fig.~\ref{fig:pdf_density}. This figure shows the effect of viscosity in the density distribution within a thin shell of 5~kpc around the virial radius of the cluster g55. The vertical dashed lines indicate the position of the density threshold to split the distribution into bulk and high-density gas. Once the gas has been separated, we fit the bulk gas to a log-normal distribution, indicated with the dash-dotted lines. Even though the mean value of the distribution does not change much, viscosity broadens the distribution. This result shows quantitatively what we observed in Fig.~\ref{fig:colormaps_density}. Viscosity produces larger density fluctuations, where particles deviate more from the mean value of the density distribution than in the non-viscous case. As a consequence, choosing a larger value of $f_{\mathrm{cut}}$ would displace the threshold in the viscous case further towards larger densities, taking particles of the high-density tail as bulk gas. On the other hand, choosing a rather low value of $f_{\mathrm{cut}}$ means that we might be underestimating the threshold in the ideal case (see appendix \ref{app:fcut}). For this reason, we take the lowest value suggested by \cite{Zhuravleva_2012}.

\subsubsection{Density fluctuations} \label{sec:density_fluct}
\begin{figure*}
    \centering
	\includegraphics[width=\textwidth]{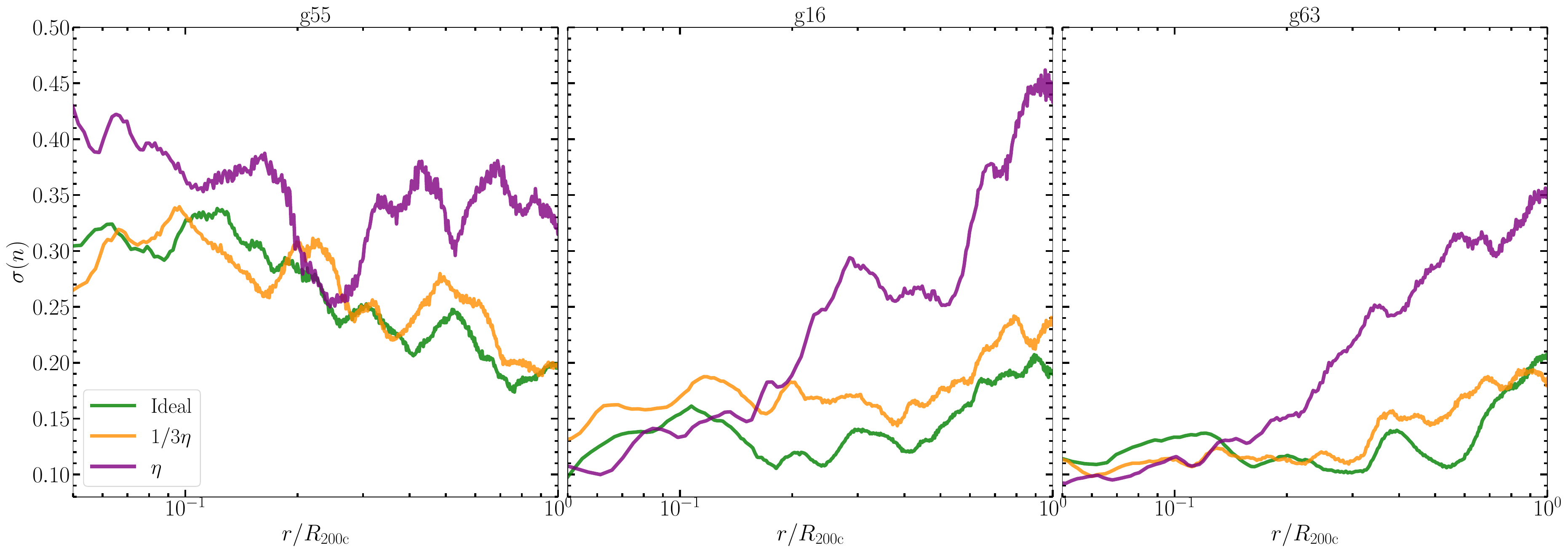}
    \caption{Radial profile of the standard deviation of the log-normal fit for each one of the spherical shells. From left to right: clusters g55, g16 and g63 for the different runs.}
    \label{fig:sigma_three}
\end{figure*}

Fig.~\ref{fig:pdf_density}, however, only shows a thin slice of cluster g55 to illustrate the effect of viscosity on the gas distribution and the method employed to separate the bulk gas from the high-density gas. To understand the overall effect of viscosity in the whole cluster, Fig.~\ref{fig:sigma_three} shows the variation of the standard deviation ($\sigma$) of the log-normal fit as a function of the radius for the three different clusters. Assuming the bulk gas follows a log-normal distribution, the value of $\sigma$ can be understood as a density fluctuation measurement (i.e. the width of the distribution), since
\begin{equation}
    \frac{\delta \rho}{\rho} = \log_{10}\frac{\rho_2}{\rho_1} = \frac{2\sqrt{2\ln 2}}{\ln 10} \, \sigma \simeq 1.02 \sigma \, , 
    \label{eqn:density_fluct}
\end{equation}
where, as before, $\rho_1$ and $\rho_2$ correspond to the 12th and the 88th percentile of the density, respectively. Increasing the amount of viscosity produces broader distributions for all the shells along the radius, which translates into larger density fluctuations. Overall, with full Spitzer viscosity we get the largest fluctuations. Comparing the case with $1/3 \, \eta$ and the ideal case we still see larger fluctuations in the runs with $1/3 \, \eta$, although the differences are not large.

\subsubsection{Density vs velocity fluctuations} \label{sec:dens_vel_fluct}
\begin{figure*}
    \centering
	\includegraphics[width=\textwidth]{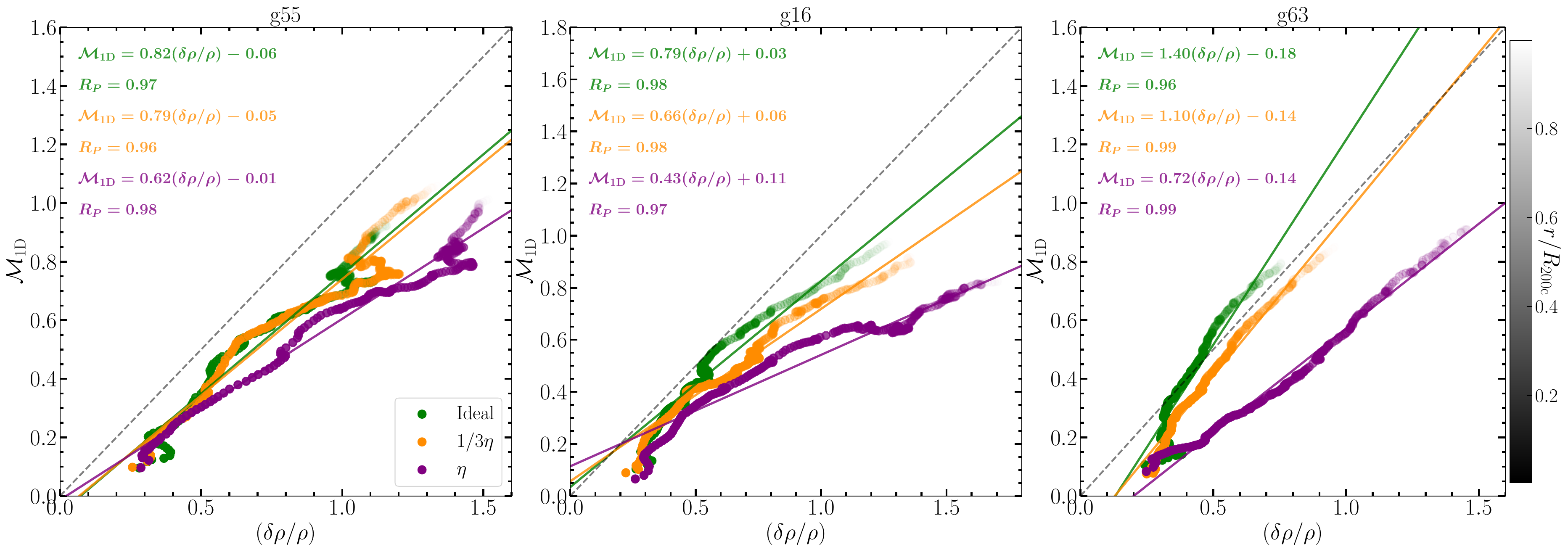}
    \caption{Velocity fluctuations against density fluctuations within $R_{\mathrm{200c}}$. The transparency of the dots indicates the distance to the center of the cluster. The dots are calculated by averaging over 40 snapshots from redshift 0.4 to redshift 0 for each radial shell. For each of the runs we do a linear fit, indicating the slope, intercept and the Pearson's coefficient for each case. From left to right: clusters g55, g16 and g63 for each one of the configurations.}
    \label{fig:dens_vs_vel_fluct}
\end{figure*}

As explained in section \ref{sec:theory}, the density fluctuations are expected to scale linearly with velocity fluctuations. In the ICM, it has been found that the density fluctuations correlate with $\mathcal{M}_{\mathrm{1D}}$ with a slope close to 1 \citep{Zhuravleva_2014, Gaspari_2014}. However, this might also depend on the dynamical state of the cluster, where unrelaxed clusters might deviate from that value \citep{Simonte_2022, Zhuravleva_2023}. We want to investigate how viscosity affects that relation. To do so, we first calculate the velocity fluctuations via the root mean square velocity ($v_{rms}$) as
\begin{equation}
    \delta v = \sqrt{(v_{\mathrm{x}} - \langle v_{\mathrm{x}}\rangle )^2 + (v_{\mathrm{y}} - \langle v_{\mathrm{y}}\rangle )^2 + (v_{\mathrm{z}} - \langle v_{\mathrm{z}}\rangle )^2} \, ,
    \label{eqn:vel_fluct}
\end{equation}
where $\langle v_{\mathrm{x}}\rangle$, $\langle v_{\mathrm{y}}\rangle$ and $\langle v_{\mathrm{z}}\rangle$ are the mean velocities in each shell. We then calculate the 1D mach number of each particle as $\mathcal{M}_{\mathrm{1D}} = \delta v / (c_s \, \sqrt{3})$. We take the region within $R_{\mathrm{200c}}$ of the cluster, similar to the one taken in \cite{Zhuravleva_2023}. For better statistics, we take the last 40 snapshots from $z\sim0.4$ to $z=0$ and calculate the mean value of density and velocity fluctuations of each shell in each snapshot using equations \ref{eqn:density_fluct} and \ref{eqn:vel_fluct} respectively. For each snapshot we take the same radial bins normalised to $R_{\mathrm{200c}}$. Finally, we take the mean value of the density and velocity fluctuations for each shell over the snapshots and plot $\mathcal{M}_{\mathrm{1D}}$ against $\delta \rho / \rho$, as shown in Fig.~\ref{fig:dens_vs_vel_fluct}. Since we have taken the average over a period of time, we cannot take into account the dynamical state of the clusters here.

There is a clear linear trend in all clusters, independently of the amount of viscosity, with a low scatter in all cases (the Pearson's correlation coefficient is close to 1, indicating a good correlation). Viscosity reduces the slopes of the linear fits\footnote{Note that the slope is the inverse $\eta_{\rho}$.}. We observe that density and velocity fluctuations increase further from the center (lighter data points). This reflects the state of the ICM in clusters that keep growing by accretion \citep{Zhuravleva_2023}. The values of $\mathcal{M}_{\mathrm{1D}}$ are consistent with direct observations of Perseus \citep{Hitomi_2018}, where they found $\mathcal{M}_{\mathrm{3D}} = \sqrt{3} \mathcal{M}_{\mathrm{1D}} \simeq 0.3-0.45$ within the central 100kpc of the cluster ($\sim 0.03R_{\mathrm{200c}}$) and indirect measurements from observations \citep{Lovisari_2024, Dupourque_2024},  with $\mathcal{M}_{\mathrm{3D}} \simeq 0.36-0.41$ for relaxed clusters in the inner regions of the cluster (within $R_{\mathrm{500c}} \sim 0.7R_{\mathrm{200c}}$). 

\begin{figure*}
    \centering
	\includegraphics[width=\textwidth]{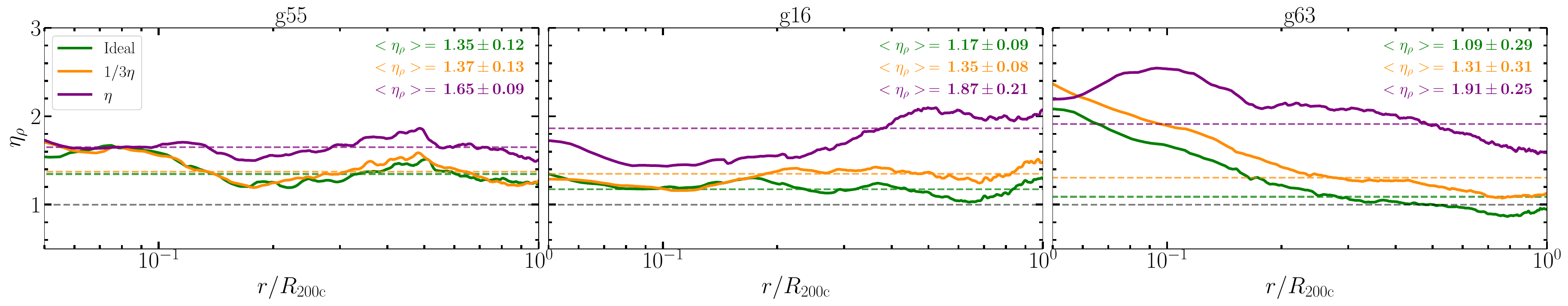}
    \caption{Ratio between density fluctuations and velocity fluctuations as a function of the radius within $R_{\mathrm{200c}}$ (same data as in Fig.~\ref{fig:dens_vs_vel_fluct}). The solid lines indicate the data from the simulations, while the dashed lines indicate the mean value in each case within $R_{\mathrm{200c}}$. The mean value and the standard deviation calculated over all the shells in each run is indicated in each panel. From left to right: clusters g55, g16 and g63 for each one of the configurations.}
    \label{fig:eta_rho_radius}
\end{figure*}

To study in more detail the ratio of density to velocity fluctuations, Fig.~\ref{fig:eta_rho_radius} shows the radial dependence of this ratio ($\eta_\rho$). 

The cases with full viscosity have consistently a larger value of $\eta_{\rho}$, as a consequence of the shallower slope seen in Fig.~\ref{fig:dens_vs_vel_fluct}. Cluster g63 shows a mean value closer to one (although a large scatter) in the inviscid case compared to clusters g55 and g16, all in agreement with the results found previously by \cite{Simonte_2022} and \cite{Zhuravleva_2023}. The values of $\eta_{\rho}$ in the cases with $1/3 \, \eta$ do not change significantly among the different clusters, whereas the runs with full viscosity have a larger value of $\eta_{\rho}$.

\subsection{Comparison with observations} \label{sec:observations}

We have discussed the effects that viscosity has in galaxy clusters, focusing so far on the comparison between viscid and inviscid simulations. However, for a comparison with observations we need to calculate the 3D amplitude fluctuations as a function of the wavenumber $k = 1 / l$ (where $l$ is the length scale of the fluctuations). To do so, we first need to calculate the density fluctuations of our data elements and then compute the power spectrum. In particle based simulations the computation of the Fourier Transform to calculate the power spectrum is not possible, so we need to interpolate the particle properties onto a grid. By doing so, we also avoid spurious effects caused by the voids after removing the high-density regions (see section \ref{sec:gas_distribution}). In grid simulations, the voids are replaced by the median value of the shell \citep[e.g.][]{Zhuravleva_2023}. In our case, the voids are automatically replaced by an average value over the particles of the region due to the interpolation method employed. This is done using the code \textsc{vortex-p} \citep[][]{Valles-Perez_2021,Valles-Perez_2021b, Valles-Perez_2024}, which creates an ad-hoc AMR mesh hierarchy from the density and velocity fields with the ultimate goal of providing a multi-resolution Helmholtz-Hodge and Reynolds decomposition. In this work, we use it exclusively to assign the density and velocity fields onto an AMR mesh, which is done using the same kernel configuration as for evolving the simulation. Once the particle data have been interpolated into the mesh, we calculate the density fluctuations. This is done by decomposing the density field into unperturbed and fluctuating components \citep{Churazov_2012}:
\begin{equation}
    \rho(x,y,z) = \rho_0(x,y,z)[1+\delta(x,y,z)] \, ,
    \label{eqn:density_distribution}
\end{equation}
where $\rho_0(x,y,z)$ is the unperturbed density distribution and 
\begin{equation}
    \delta(x,y,z) = \frac{\delta \rho}{\rho_0} = \frac{|\rho_i - \rho_0|}{\rho_0} = \frac{|\rho_i - \langle \rho \rangle|}{\langle \rho \rangle}
    \label{eqn:density_fluctuations}
\end{equation}
are the fluctuations. $\rho_i$ is the density of each cell and $\langle \rho \rangle$ is the value of the density profile at that distance from the center (see density profiles in appendix \ref{app:density_profile}). In our case, we evaluate the density profile at a given radius to estimate the unperturbed $\rho_0(x,y,z)$. For the power spectrum we assume an isotropic fluctuation field that is a function of $k = \sqrt{k_x^2 + k_y^2 + k_z^2}$. The 3D amplitude fluctuations can be computed from the power spectrum as
\begin{equation}
    \left( \frac{\delta \rho}{\rho} \right)_k = \sqrt{4\pi P(k) k^3} \, ,
    \label{eqn:3d_fluctuations}
\end{equation}
where $P(k)$ is the power spectrum. 

\begin{figure*}
    \centering
	\includegraphics[width=\textwidth]{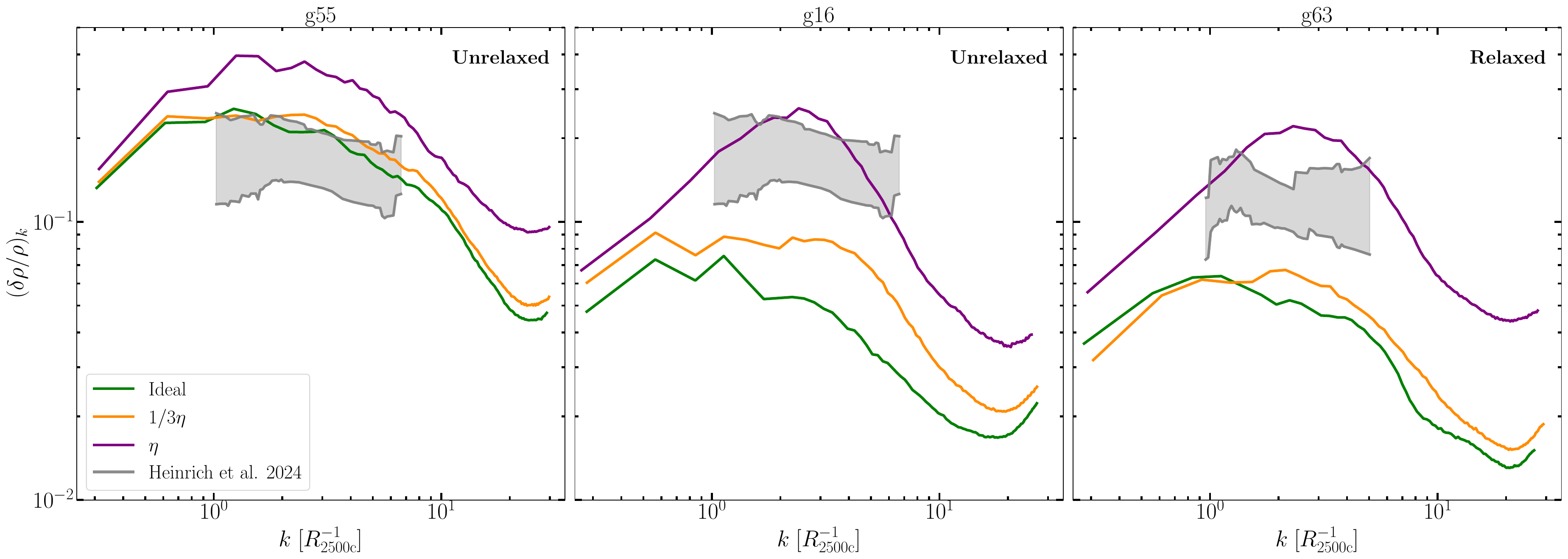}
    \caption{Comparison between density fluctuations obtained from our simulations with different viscosities (solid lines) and observations from \cite{Heinrich_2024} (grey areas) as a function of the scale within $R_{\mathrm{2500c}}$. From left to right: clusters g55, g16 and g63 for each one of the configurations. The unrelaxed clusters (g55 and g16) are compared with the observational data of unrelaxed clusters and the relaxed cluster (g63) with the observational data of relaxed clusters.}
    \label{fig:rho_spectrum}
\end{figure*}

These 3D amplitude density fluctuations can be observed from X-ray emission and can be used to constrain the amount of viscosity in the ICM \citep{Zhuravleva_2019}. Fig.~\ref{fig:rho_spectrum} shows a comparison of density fluctuations of our sample of clusters for both unrelaxed (g55 and g16) and relaxed clusters (g63), with observations taken from \cite{Heinrich_2024}. The observational data is composed by 80 clusters (24 relaxed, 30 intermediate and 26 unrelaxed) of $M_{\mathrm{2500c}} \in [8\times10^{13}, 10^{15}]$~M$_{\mathrm{\odot}}$, comparable to our sample (our clusters have $M_{\mathrm{2500c}} \sim 2.4 - 3.9\times10^{14} $M$_{\odot}$). For a better comparison with observations, we take the same region of the cluster as in \citet{Heinrich_2024}, i.e. the gas inside $R_{\mathrm{2500c}}$. The grey region indicates the average at each scale of the density fluctuations for the relaxed (unrelaxed) subsample plus/minus one standard deviation. 

In all cases, viscosity leads to a larger amplitude of the fluctuations at all scales, leading to around two times larger amplitude in the cases with full viscosity. Furthermore, the case with $1/3 \, \eta$ has similar results compared to the non-viscous case in g55 and g63, but it is $\sim30\%$ higher in g16. Although in cluster g55 the $1/3 \, \eta$ and non-viscous runs are more consistent with observations, in clusters g16 and g63 the full viscosity case matches better the overall amplitude of observations. However, due to the small sample of simulations that we have, it is difficult to exclude any particular amount of viscosity by comparing our results with the overall amplitude of the observations.

To constrain the amount of viscosity in the ICM, \cite{Zhuravleva_2019} suggested a criterion based on studying the slope of density fluctuations spectrum. In the cases with higher viscosity the turbulent cascade is stopped earlier, leading to a steeper power spectrum that translates into steeper 3D density amplitude fluctuations. This was probed by comparing Coma observations with direct numerical simulations (DNS) of hydrodynamic turbulence. They found a level of suppression of at least a factor of 10 with respect to the Spitzer value, depending on the Prandtl number\footnote{The Prandtl number is defined as the ratio of momentum to thermal diffusivity (i.e. the ratio between viscosity and thermal conduction).}.

Assuming that the motions are subsonic and driven at large scales, one can assume that the fluctuations in the ICM are passively advected by the velocity field \citep{Kunz_2022}. Therefore, a comparison with hydrodynamic turbulence simulation is somewhat reasonable. However, the DNS simulations employed in \cite{Zhuravleva_2019} use a rather idealised setup, where compressible fluids (important mainly for unrelaxed clusters) or magnetic fields are not considered. They also assumed an isothermal fluid of $T_e \sim 8$~keV ($\sim9.28\times10^7$~K) for Coma. And, since viscosity is highly dependent on the temperature (see equation \ref{eqn:viscosity}), that means that they used a constant value of viscosity. This might produce the sharp viscous cutoff observed in their results. However, in cosmological simulations, we observe a range of temperatures from $3.3\times10^6$~K to $2.5\times10^8$~K, and therefore, a big range of values for viscosity. This range of values for viscosity at different scales prevents the power spectrum from having a sharp cutoff, as observed in Fig.~\ref{fig:rho_spectrum} for density and Fig.~\ref{fig:velocity_power_spectrum} for velocities power spectrum. As a consequence, the slope of the 3D amplitude of density fluctuations is insensitive to the amount of viscosity. This is investigated in more detail in section \ref{sec:thermodynamics}.

It is important to note the maximum observed at $k \sim 2.5 R_{\mathrm{2500c}}^{-1} \sim 175$~kpc in the viscous runs of clusters g16 and g63 in Fig.~\ref{fig:rho_spectrum}. This feature is not found either in observations or the other simulations, where the amplitude of density fluctuations decreases monotonically with $k$. The reason is the value of $f_{\mathrm{cut}}$ used in equation \ref{eqn:threshold}, where we used a value of $f_{\mathrm{cut}} = 2.5$, the minimum value suggested in \cite{Zhuravleva_2012}. This value was reasonable for our cluster g55, where we have a clear separation of bulk gas following a log-normal distribution and a high-density tail (see Fig.~\ref{fig:pdf_density}). However, there is no clear separation between the bulk gas and high-density regions in the density PDF of cluster g16 (see appendix \ref{app:fcut}) and g63, which looks more like a log-normal plus power law distribution. Due to its broad PDF, the median value is large and a value of $f_{\mathrm{cut}} = 2.5$ sets a threshold at a large density value, causing very few or no particles to be considered as high-density (see Fig.~\ref{fig:pdf_g16}). The small fraction of particles removed due to the choice of $f_{\mathrm{cut}}$ increases the amplitude density fluctuations at the scale of the high-density clumps ($50$~kpc - $200$~kpc). A more reasonable value for this cluster would be $f_{\mathrm{cut}} = 0.5$, where the split between bulk and high-density gas is more appropriate, as shown in the top panel of Fig.~\ref{fig:factor_pdf}. With a lower value of $f_{\mathrm{cut}}$, the maximum observed in the middle panel of Fig.~\ref{fig:rho_spectrum} disappears, as can be seen in Fig.~\ref{fig:factor_specrtum}. However, the same value should be applied to all the analysis for consistency, and that would imply that in cluster g55 a lot of bulk gas would be considered as high-density gas (see bottom panel of Fig.~\ref{fig:factor_pdf}). For this reason, we perform our analysis with a value of $f_{\mathrm{cut}} = 2.5$, acknowledging that it has an impact on the viscous runs of clusters g16 and g63, leading to a factor of $\sim 2$ larger amplitude at the scales of the high-density clumps ($\sim 175$~kpc). However, the slope of the density fluctuations is not affected by the choice of $f_{\mathrm{cut}}$ (see Fig.~\ref{fig:factor_specrtum}), indicating that the method to constrain viscosity is still robust regardless of the value of $f_{\mathrm{cut}}$.

\begin{figure*}
    \centering
	\includegraphics[width=\textwidth]{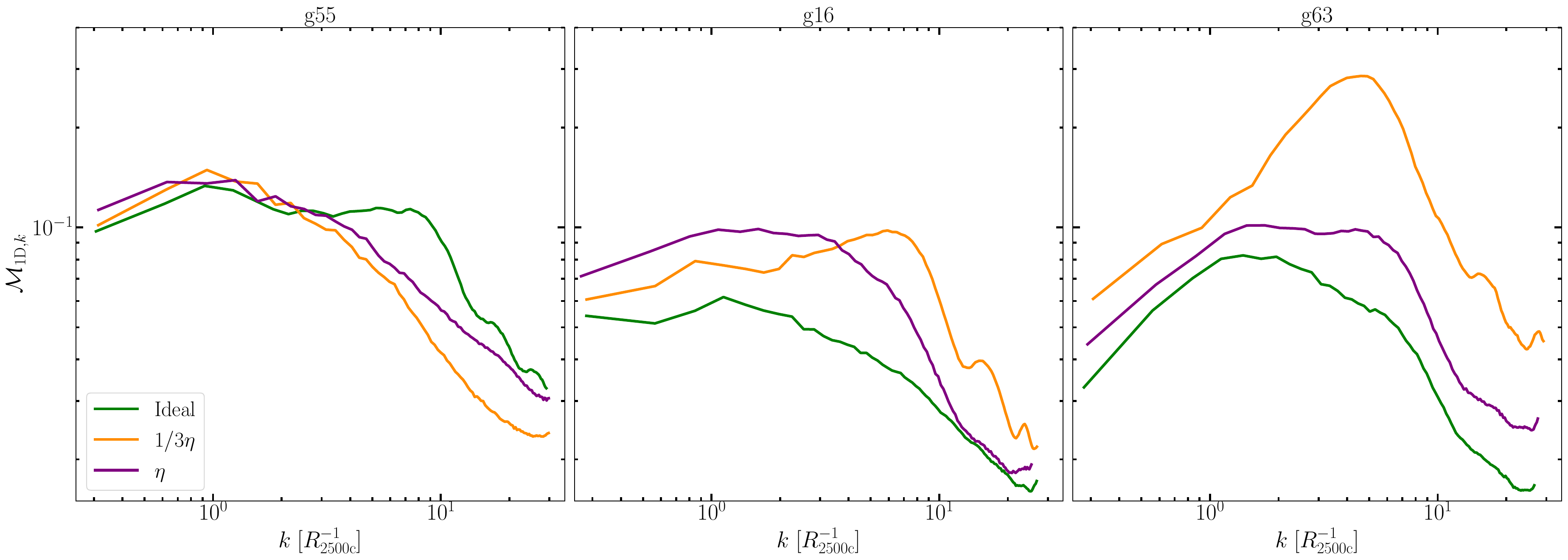}
    \caption{Comparison of velocity fluctuations obtained from our simulations with different viscosities as function of the scale within $R_{\mathrm{2500c}}$. From left to right: clusters g55, g16 and g63 for each one of the configurations.}
    \label{fig:vel_fluct}
\end{figure*}

The code \textsc{vortex-p} also allows us to interpolate the velocity fluctuation field calculated from $v_{rms}$ as shown in equation \ref{eqn:vel_fluct} and the soundspeed of each particle. Then we calculate the power spectrum of the velocity fluctuations in an analogous way as shown in equation \ref{eqn:3d_fluctuations}
\begin{equation}
    \left( \frac{\delta v}{c_s} \right)_{k} = \sqrt{4\pi P_{\mathrm{3D}}(k) k^3} = \mathcal{M}_{\mathrm{3D}, k} \, ,
    \label{eqn:velocity_fluct}
\end{equation}
where $P_{\mathrm{3D}}(k) = P_x(k) + P_y(k) + P_z(k)$. To obtain the 1D velocity fluctuations we do $\mathcal{M}_{\mathrm{1D},k} = \mathcal{M}_{\mathrm{3D},k} / \sqrt{3}$. Fig.~\ref{fig:vel_fluct} shows the velocity fluctuations for the three different clusters. 

Since viscosity reduces the velocity of the particles \citep{Marin-Gilabert_2022} and, at the same time, increases the temperature (see section \ref{sec:morphology}) and, therefore, the soundspeed, one could expect that the $\mathcal{M}_{\mathrm{1D},k}$ decreases with viscosity. However, Fig.~\ref{fig:vel_fluct} shows very similar results among the simulations or even larger amplitude fluctuations in the viscous cases. This apparently counter-intuitive result might be due to the lack of mixing in the viscosity runs. This causes that the most massive structures in the cluster keep their mass for a longer time compered to the non-viscous runs, experiencing a larger gravitational pull and accelerating as they fall into the gravitational well of the cluster before being stripped. This can be observed also in the kinetic energy spectrum in appendix \ref{app:vel_power_spectrum}, where at large scales the runs with viscosity lead to larger values of the energy spectrum.

\begin{figure*}
    \centering
	\includegraphics[width=\textwidth]{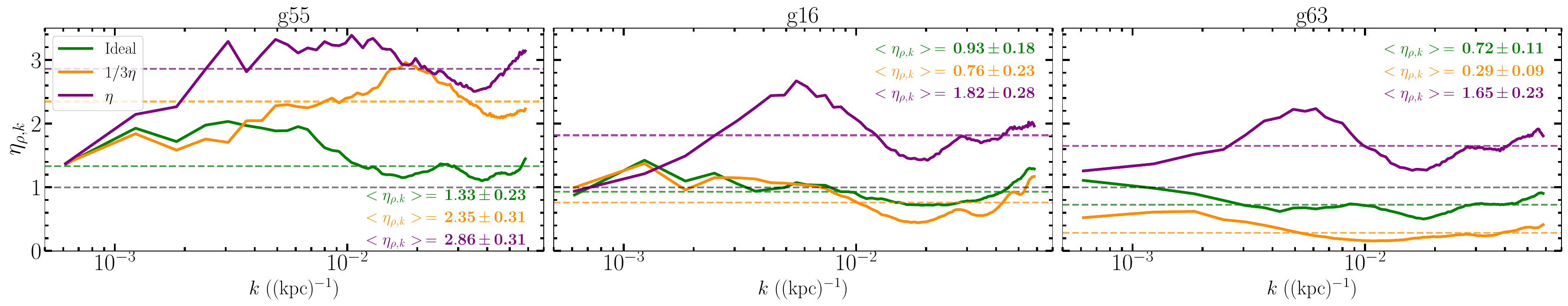}
    \caption{Ratio between density fluctuations and velocity fluctuations as function of the scale within $R_{\mathrm{2500c}}$. The solid lines indicate the data from the simulations, while the dashed lines indicate the mean value in each case along the different scales. The mean value and the standard deviation in each run is indicated in each panel. From left to right: clusters g55, g16 and g63 for each one of the configurations.}
    \label{fig:eta_spectrum}
\end{figure*}

Considering density and velocity fluctuations, we can study the relationship between the density and velocity fluctuations of equation \ref{eqn:eta_relation}. In Fig.~\ref{fig:eta_spectrum} we show the value of $\eta_{\rho, k}$ calculated as $\eta_{\rho, k} = (\delta \rho / \rho)_k / \mathcal{M}_{\mathrm{1D},k}$ for different scales. The runs with greater viscosity tend to have larger values of $\eta_{\rho, k}$. This means that, even though both density and velocity fluctuations are higher in the viscous runs, they do not compensate for each other and the ratio between the two is still larger than for the non-viscous runs. The cluster g55 has values larger than 1, however this is in agreement with previous work for unrelaxed clusters \citep{Gaspari_2014, Simonte_2022}. The cluster g16 shows a relation very close to 1, as expected theoretically \citep{Zhuravleva_2014,Zhuravleva_2023}, although in the case with full viscosity we still see the maximum analysed in appendix \ref{app:fcut}. In cluster g63 we observe a value close to, but lower than 1, also consistent with previous work for a very relaxed cluster.

\subsection{Thermodynamical structure} \label{sec:thermodynamics}

\begin{figure*}
    \centering
	\includegraphics[width=\textwidth]{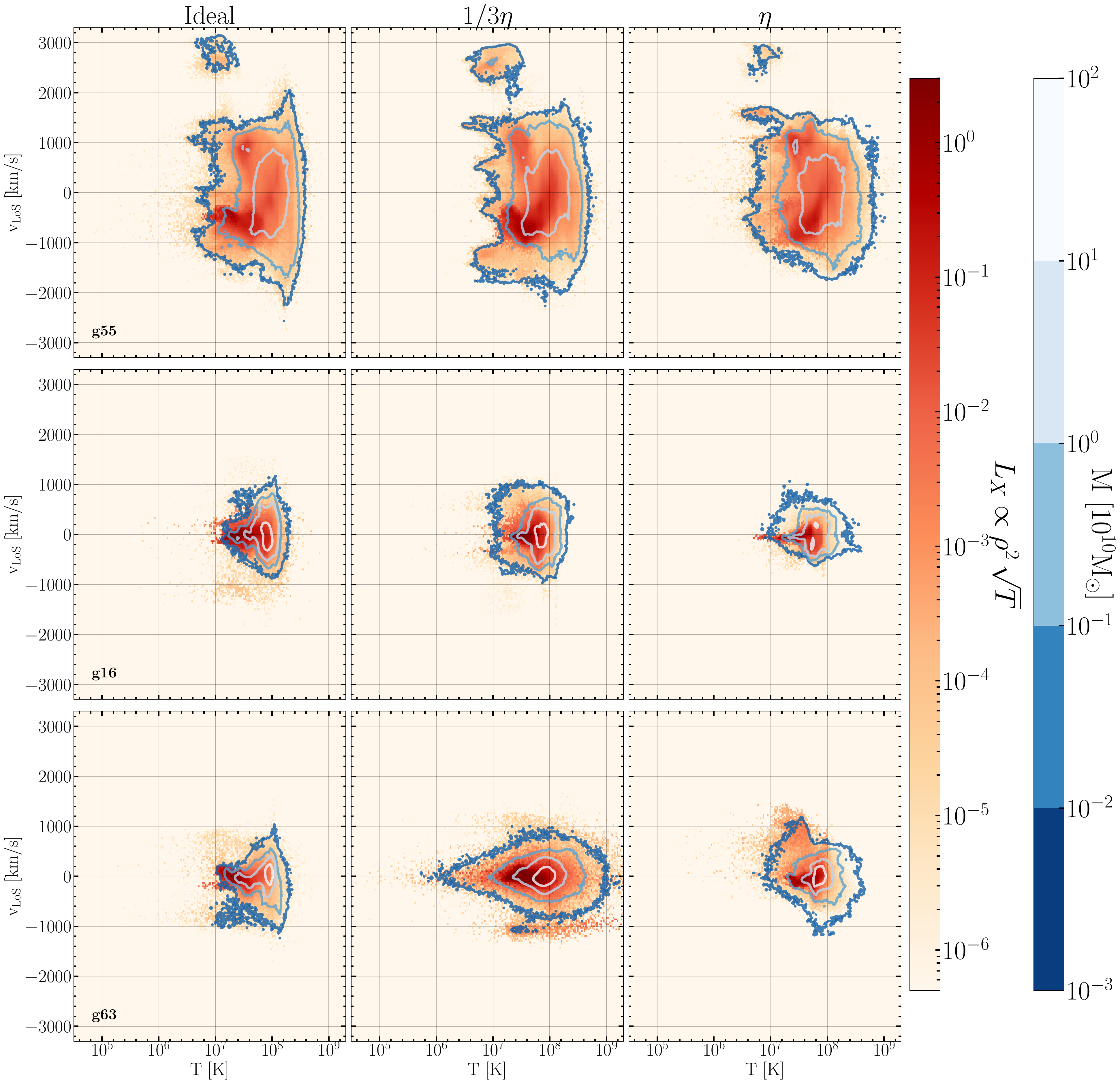}
    \caption{2D histograms of the $v_{\mathrm{LoS}}$ as a function of the temperature, color-coded by emissivity. The contours indicate the mass distribution of the particles within the cluster. From left to right: MHD only, MHD with $1/3$ of Spitzer viscosity and full Spitzer viscosity at redshift zero. From top to bottom: clusters g55, g16 and g63.}
    \label{fig:2D_hist_T_vel}
\end{figure*}

For a better understanding of the gas dynamics of the clusters, in this section, we study the thermodynamical structure of the different clusters. For a better comparison with observations, we focus only on the inner regions of the cluster (within $R_{\mathrm{2500c}}$), as we did in previous sections. To visualize the thermodynamical structure of the clusters, Fig.~\ref{fig:2D_hist_T_vel} shows 2D histograms of the line-of-sight velocity ($v_{\mathrm{LoS}}$) as a function of the temperature, color-coded by the emissivity. For the calculation of the emissivity, we assume it to be $\propto \rho^2 \sqrt{T}$, expected for thermal bremsstrahlung \citep{Sarazin_1986}. We also overplot the mass distribution of the particles as a contour for the different $v_{\mathrm{LoS}}$ and temperatures.

The unrelaxed dynamical state of cluster g55 (top row) can be identified by analysing the $v_{\mathrm{LoS}}$, where the mass is distributed over a larger range of velocities. The emissivity is also distributed over a larger range of velocities and there is not a clear emissivity peak, as can be seen in the other two clusters. Although cluster g16 is unrelaxed and g63 is relaxed, the distributions of $v_{\mathrm{LoS}}$ look very similar, being approximately $v_{\mathrm{LoS}} \in [-1000, 1000]$~km/s in both cases. The peak of emissivity in both g16 and g63 is localised around $v_{\mathrm{LoS}} = 0$~km/s for all the configurations in the two clusters.

In all cases, the peak of emissivity is approximately between $10^7$ and $10^8$~K, within the range expected for galaxy clusters \citep{Biffi_2012}. However, cluster g55 exhibits a broader temperature distribution due to its unrelaxed state, reaching temperatures of $5\times10^8$~K. The cases with full viscosity tend to have more mass at larger temperatures, as can be seen also in Fig.~\ref{fig:colormaps_temperature}.

It is important to note that cluster g16 exhibits a narrower mass-weighted temperature distribution in the three configurations. The majority of the mass is concentrated in a narrow range of temperatures, meaning that this cluster is closer to being isothermal for the three runs than the other two clusters. This is similar to the case analysed by \citet{Zhuravleva_2019}, where they assumed an isothermal fluid of $T \sim 9.28\times10^7$~K for Coma. The narrower the temperature distribution is, the narrower is the range of viscosities as well (see equation \ref{eqn:viscosity}). This would imply a sharper cutoff in the turbulent cascade, leading to a steeper 3D amplitude density fluctuations. Although cluster g16 is not perfectly isothermal, these steeper 3D amplitude density fluctuations can be seen in the middle panel of Fig.~\ref{fig:rho_spectrum} by comparing the non-viscous and the viscous case. The maximum produced by the choice of $f_{\mathrm{cut}}$ might affect the results, however the steeper spectrum can also be seen in Fig.~\ref{fig:factor_specrtum} for different values of $f_{\mathrm{cut}}$. The difference in slope is not as prominent as the one suggested by \citet{Zhuravleva_2019}, however, it indicates that their criterion to constrain viscosity might be applicable only to isothermal galaxy clusters. The consequence of the narrower temperature distribution can also be seen in the middle panel of Fig.~\ref{fig:velocity_power_spectrum}, where the dynamical range of the spectrum reaches smaller scales in the non-viscous than in the viscous case. 

Fig.~\ref{fig:2D_hist_T_vel} shows how a more isothermal cluster produces a steeper slope in the density fluctuations amplitude for the cases with viscosity compared to the inviscid cases. However, the differences in the slope vanish if the temperature distribution within the cluster is broader.

\section{Conclusions} \label{sec:conclusions}

In this work we performed a set of cosmological simulations using the code \textsc{OpenGadget3} of three different galaxy clusters, each with three different values of viscosity to quantify the effect of viscosity in the ICM. First, we compared the results of these simulations at redshift zero in order to understand the effect that viscosity has in the ICM. Then we tried to constrain the amount of viscosity in the ICM by comparing with X-ray observations of density fluctuations. Our key conclusions are:
\begin{itemize}
\item Although the overall morphology remains the same, by visual inspection one can identify morphological differences produced by viscosity due the suppression of instabilities on small scales. The runs with more viscosity show a larger amount of small clumps that have not been disrupted by instabilities; a more filamentary structure produced by the gas stripped from infalling structures towards the center of the cluster; and more gas concentrated in the denser regions, rather than the more mixed and homogeneous gas seen in the non-viscous cases.

\item The kinetic energy transformed into internal energy by viscosity leads to the heating of the less dense regions, although the denser regions remain as cold as in the inviscid runs. This is translated into a higher virial temperature in the runs with viscosity by $\sim$5\% - 10\%. 

\item The lack of mixing in the viscous case produces a broader density PDF of the bulk gas of the cluster, which can be interpreted as larger density fluctuations. The density fluctuations are consistently larger along the cluster radius the more viscous the medium is, increasing towards longer distances from the center. 

\item Using a fixed value of $f_{\mathrm{cut}}$, used to divide the bulk and high-density gas, works well for inviscid simulations of galaxy clusters. However, the results with full viscosity (where the density PDF is broader) are slightly dependent on the choice of $f_{\mathrm{cut}}$. Therefore, a more accurate method to separate bulk and high-density gas should be investigated in the future to avoid the dependence on the shape of the density PDF.

\item The density and velocity fluctuations are directly proportional to one another and both increase with the distance from the center. However, in all three clusters the slope of the relationship between velocity and density fluctuations decreases with increasing viscosity. This linear relation is translated into a density to velocity fluctuations ratio of the order of unity for distances up to $R_{\mathrm{200c}}$. 

\item The runs with viscosity tend to have larger amplitude of density fluctuations as a function of the scale. However, the density fluctuations obtained from our simulations are consistent with observations, even in the case with full viscosity. This is due to the large range of temperature distribution. In isothermal clusters, the slope of the amplitude of density fluctuations is affected by viscosity. This means that the method suggested in \citet{Zhuravleva_2019} to constrain the amount of viscosity in the ICM by measuring the slope of the amplitude of density fluctuations is only applicable to isothermal clusters. 

\item The amplitude of velocity fluctuations also appears to increase with viscosity, mainly at large scales. This behaviour can also be observed when computing the velocity power spectrum, where viscosity leads to more power at large scales, but decreases at smaller scales. The ratio of density to velocity fluctuations tends to be higher for cases with full viscosity, while remaining close to unity for all cases regardless of the amount of viscosity.

\end{itemize}

In summary, the cosmological simulations with viscosity show some morphological differences as well as temperature differences. We can quantify these morphological differences by measuring the fluctuations in density, which become larger in the cases with higher viscosity. This can be compared with observational measurements of the amplitude of density fluctuations. Our results are consistent with observations, making difficult the task of constraining the amount of viscosity in the ICM solely from density fluctuations. The velocity fluctuations also happen to be larger at large scales in the cases with viscosity, although the ratio of density to velocity fluctuations is larger with viscosity, but close to one. 

Due to the computational costs of running cosmological simulations at high resolution including physical viscosity, we could only compare three different galaxy clusters. Future work with more clusters could help us in doing statistics and not relying our results in three clusters only. Higher resolution simulations could also help us to understand better the results and avoid spurious resolution effects. Additionally, the effect of a more realistic anisotropic Braginskii viscosity cosmological simulation will be explored in future studies.

\section*{Data Availability}
The data will be made available based on reasonable request to the corresponding author.

\section*{Acknowledgments}
TM would like to thank Irina Zhuravleva and Eugene Churazov for the intense discussions which motivated this work. The authors also want to thank the referee for their very useful comments. KD and TM acknowledge support by the COMPLEX project from the European Research Council (ERC) under the European Union’s Horizon 2020 research and innovation program grant agreement ERC-2019-AdG 882679. UPS is supported by the Simons Foundation through a Flatiron Research Fellowship (FRF) a the Center for Computational Astrophysics (CCA). The CCA is supported by the Simons Foundation. MV is supported by the Italian Research Center on High Performance Computing, Big Data and Quantum Computing (ICSC), project funded by European Union - NextGenerationEU - and National Recovery and Resilience Plan (NRRP) - Mission 4 Component 2, within the activities of Spoke 3, Astrophysics and Cosmos Observations, and by the INFN Indark Grant. DVP has been supported by the Agencia Estatal de Investigación Española (AEI; grant PID2022-138855NB-C33), by the Ministerio de Ciencia e Innovación (MCIN) within the Plan de Recuperación, Transformación y Resiliencia del Gobierno de España through the project ASFAE/2022/001, with funding from European Union NextGenerationEU (PRTR-C17.I1), by the Generalitat Valenciana (grant CIPROM/2022/49), and by Universitat de València through an Atracció de Talent fellowship. This work has been supported by the Munich Excellence Cluster Origins funded by the Deutsche Forschungsgemeinschaft (DFG, German Research Foundation) under Germany’s Excellence Strategy EXC-2094 390783311. The simulations were performed at the Computational Center for Particle and Astrophysics (C2PAP).

\appendix

\section{Dependence on \lowercase{$f_{\mathrm{cut}}$}} \label{app:fcut}

\begin{figure}
    \centering
    \includegraphics[width=\columnwidth]{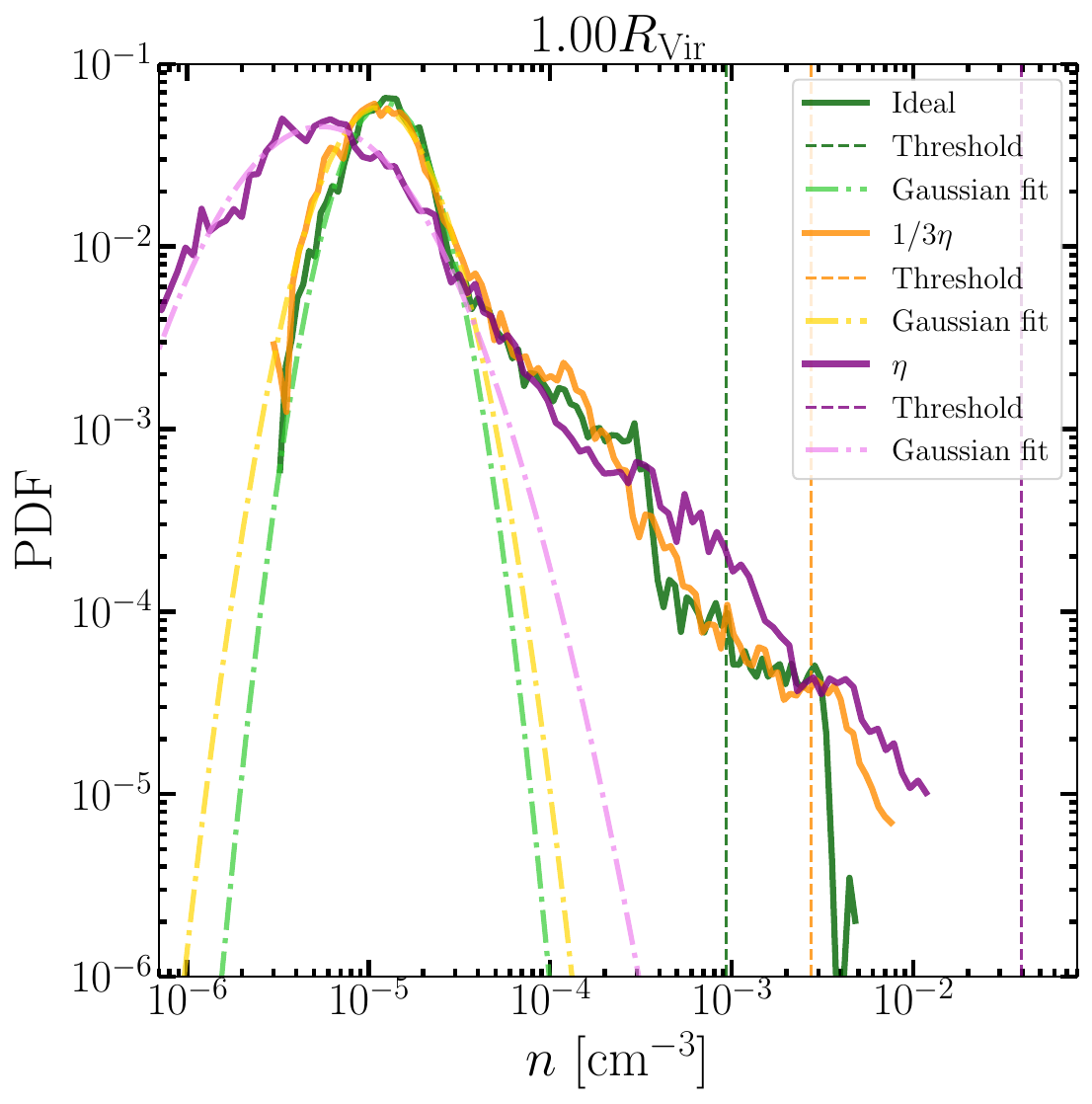}
    \caption{Density PDF of a 5kpc shell centered around the virial radius of cluster g16 using a value of $f_{\mathrm{cut}} = 2.5$. The solid lines show the data obtained from the simulations; the dashed lines the calculated threshold to split bulk and high-density gas; and the dashed-dotted lines the fit of the bulk gas to a log-normal distribution.}
    \label{fig:pdf_g16}
\end{figure}

Viscosity broadens the density PDF due to the lack of mixing. As a result, it becomes difficult to define a dividing line between the bulk and the high-density gas. The median value of density is displaced towards larger values of density, causing the threshold defined in equation \ref{eqn:threshold} to be a poor choice for the division between bulk and high-density gas. In the most extreme cases, this threshold lies outside the range of densities for certain shells of the cluster, as can be seen in Fig.~\ref{fig:pdf_g16} for cluster g16 for $f_{\mathrm{cut}} = 2.5$. 

\begin{figure}
    \centering
    \includegraphics[width=\columnwidth]{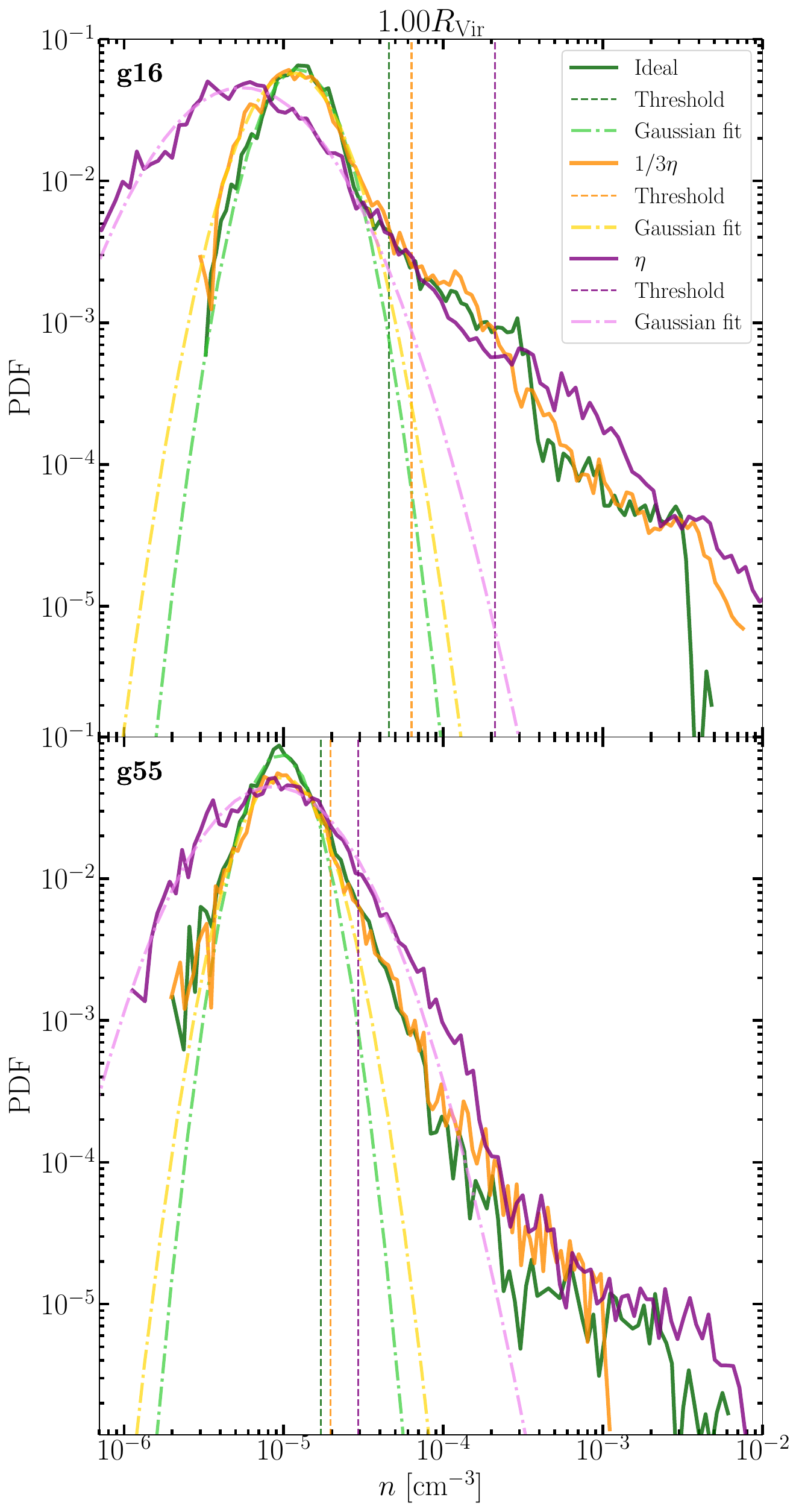}
    \caption{Density PDF of a 5kpc shell centered around the virial radius using a value of $f_{\mathrm{cut}} = 0.5$. The solid lines show the data obtained from the simulations; the dashed lines the calculated threshold to split bulk and high-density gas; and the dashed-dotted line the fit of the bulk gas to a log-normal distribution. \textit{Top panel}: cluster g16. \textit{Bottom panel}: cluster g55.}
    \label{fig:factor_pdf}
\end{figure}

\begin{figure}
    \centering
    \includegraphics[width=\columnwidth]{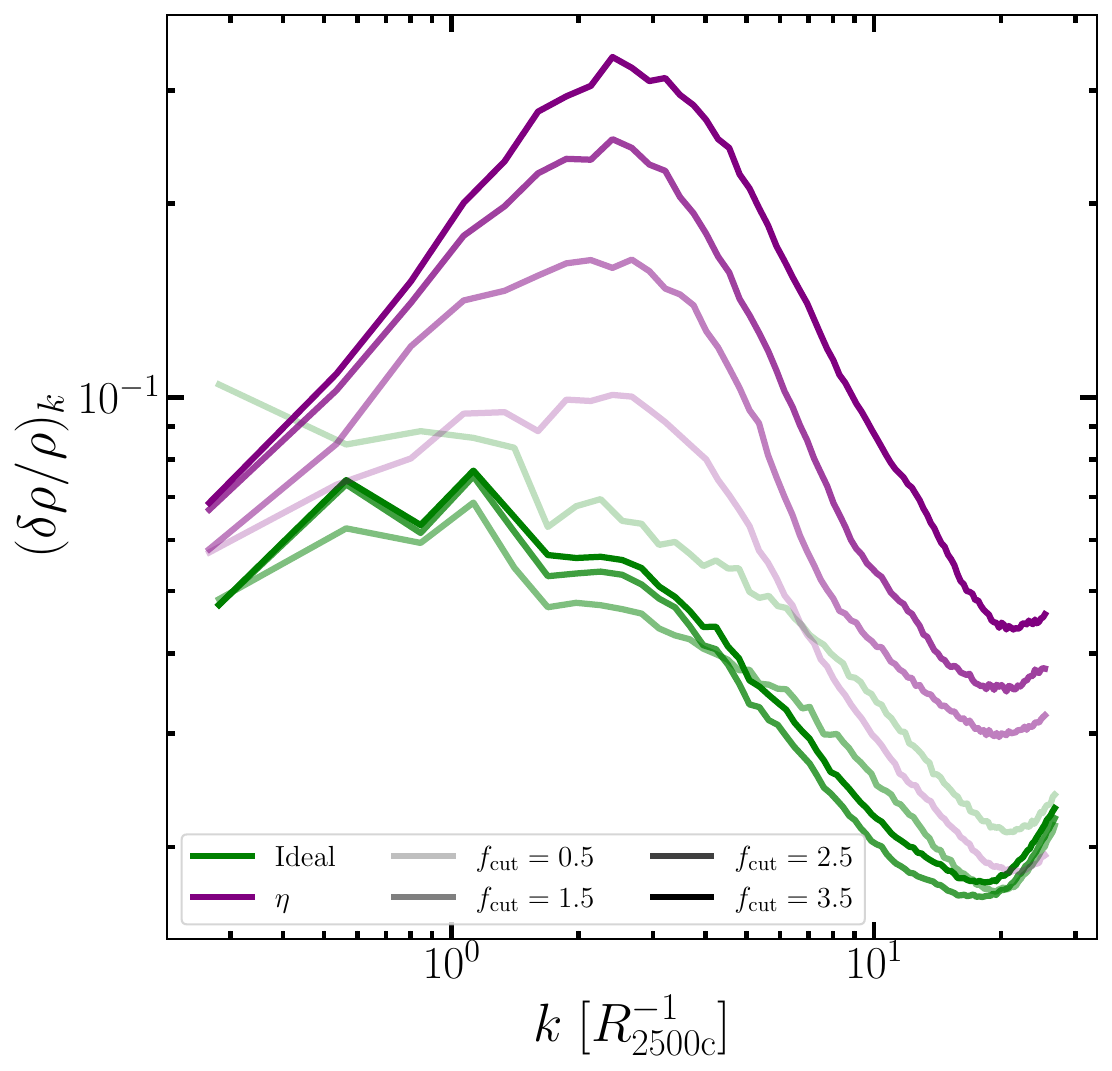}
    \caption{Comparison of the amplitude of density fluctuations in cluster g16 for different values of $f_{\mathrm{cut}}$ with and without viscosity.}
    \label{fig:factor_specrtum}
\end{figure}

This signifies that some high-density clumps are not removed, leading to the maximum in the density fluctuations spectrum, as shown in Fig.~\ref{fig:rho_spectrum}. This can be solved by reducing the value of $f_{\mathrm{cut}}$ in equation \ref{eqn:threshold} to lower values than $2.5$. By doing this, we can successfully split bulk and high-density gas, as can be seen in the upper panel of Fig.~\ref{fig:factor_pdf}, where we used $f_{\mathrm{cut}} = 0.5$. As a consequence, the maximum observed in the amplitude of the density fluctuations is reduced (see Fig.~\ref{fig:factor_specrtum}). However, if we use $f_{\mathrm{cut}} = 0.5$ in other clusters, the split of bulk and high-density gas is not properly done (see bottom panel of Fig.~\ref{fig:factor_pdf} to observe the effect in cluster g55). This value of $f_{\mathrm{cut}}$ would consider part of the bulk gas as high-density gas and would remove more gas than only high-density clumps. For consistency we use the same value of $f_{\mathrm{cut}}$ in all our clusters, choosing the value suggested in \citet{Zhuravleva_2012} of $f_{\mathrm{cut}} = 2.5$ and acknowledging the spurious effects that it can have in the amplitude of density fluctuations.

\section{Density profile} \label{app:density_profile}

\begin{figure*}
    \centering
	\includegraphics[width=\textwidth]{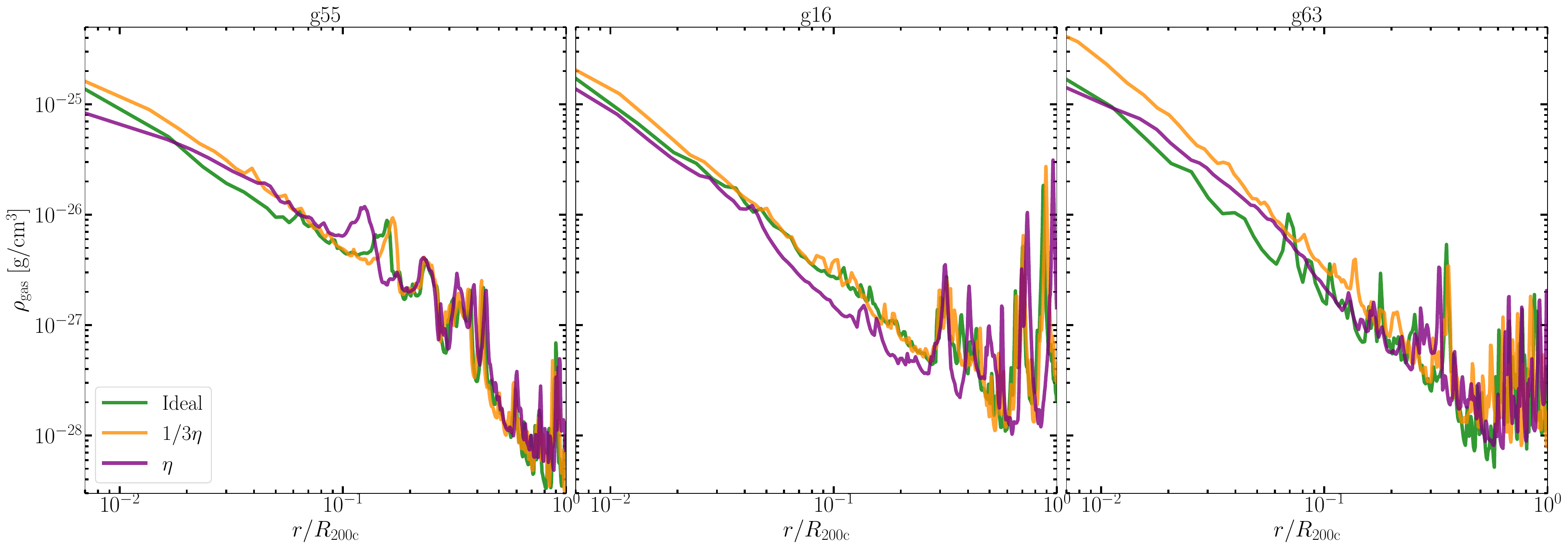}
    \caption{Radial profile of the density computed from spherical shells of 10kpc. From left to right: clusters g55, g16 and g63 for each one of the configurations.}
    \label{fig:density_profile}
\end{figure*}

If we look at the radial density profile of each one of the clusters (Fig.~\ref{fig:density_profile}) we do not observe any clear trend among the cases with different viscosities. This shows how at large scales the viscosity effects do not significantly affect the cluster morphology. We need to study the density fluctuations ($\sim 5$\%-$10$\% of the total density \citep{Churazov_2012, Sanders_2012}) to see bigger differences between the cases with and without viscosity. When making a radial profile, using the mean density for each shell hides the main impact of the viscosity, which is seen in the fluctuations about the mean. Fig.~\ref{fig:pdf_density} shows how, even though the density distribution is broader with viscosity, the mean value remains more or less the same, leading to very similar radial density profiles.

\section{Velocity power spectrum} \label{app:vel_power_spectrum}

\begin{figure*}
    \centering
	\includegraphics[width=\textwidth]{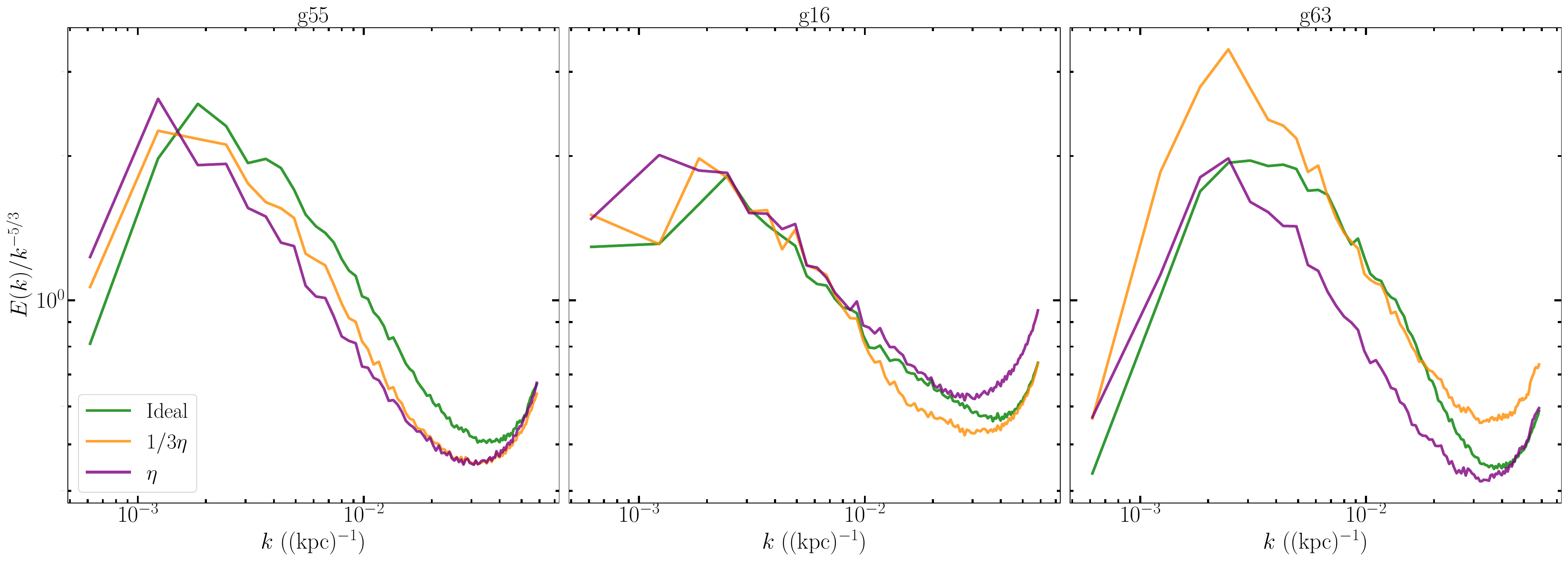}
    \caption{Velocity power spectrum normalised to Kolmogorov. From left to right: clusters g55, g16 and g63 for each one of the configurations.}
    \label{fig:velocity_power_spectrum}
\end{figure*}

We can make use of the code \textsc{vortex-p} to interpolate the particles into a grid and compute the velocity power spectrum for each cluster and for each amount of viscosity in each case. Fig.~\ref{fig:velocity_power_spectrum} shows the power spectrum normalised to Kolmogorov for each cluster within $R_{\mathrm{2500c}}$. In all cases the slope is steeper than Kolmogorov, however, the slope is the same in each cluster regardless the amount of viscosity. At larger scales the runs with viscosity appear to be more energetic, although at intermediate and small scales the viscosity runs become less energetic (except for the cluster g16).

\bibliography{sample631}{}
\bibliographystyle{aasjournal}

\end{document}